%% file: main.tex
\documentclass[pdflatex,sn-basic]{sn-jnl}

\usepackage{graphicx}%
\usepackage{multirow}%
\usepackage{amsmath,amssymb,amsfonts}%
\usepackage{amsthm}%
\usepackage{mathrsfs}%
\usepackage[title]{appendix}%
\usepackage[table]{xcolor} 
\usepackage{colortbl}      
\usepackage{textcomp}%
\usepackage{manyfoot}%
\usepackage{booktabs}%
\usepackage{algorithm}%
\usepackage{algorithmicx}%
\usepackage{algpseudocode}%
\usepackage{listings}%
\usepackage{float}   
\usepackage{caption}
\usepackage{subcaption}
\usepackage{hyperref}
\usepackage{pdflscape} 
\usepackage{tabularx}
\usepackage{microtype}
\usepackage{tocloft}

\makeatletter
\@ifundefined{Affilfont}{}{%
}
\@ifundefined{affilsep}{}{%
  \setlength{\affilsep}{1.0em}%
}
\makeatother

\theoremstyle{thmstyleone}%
\theoremstyle{thmstyletwo}%

\theoremstyle{thmstylethree}%

\begin{document}


\title[Article Title]{\textbf{Needles in a haystack: using forensic network science to uncover insider trading}}


\author*[1]{\fnm{Gian} \sur{Jaeger}}\email{gian.jaeger@oii.ox.ac.uk}

\author[2]{\fnm{Wang Ngai} \sur{Yeung}}\email{j.yeung@northeastern.edu}

\author[3]{\fnm{Renaud} \sur{Lambiotte}}\email{renaud.lambiotte@maths.ox.ac.uk}

\affil*[1]{\orgdiv{Oxford Internet Institute}, \orgname{University of Oxford}, \orgaddress{\city{Oxford}, \country{UK}}}

\affil[2]{\orgdiv{Complex Connections Lab}, \orgname{Network Science Institute}, \orgaddress{\state{London}, \country{UK}}}

\affil[3]{\orgdiv{Mathematical Institute}, \orgname{University of Oxford}, \orgaddress{\city{Oxford}, \country{UK}}}


\abstract{Although the automation and digitisation of anti-financial crime investigation has made significant progress in recent years, detecting insider trading remains a unique challenge, partly due to the limited availability of labelled data. To address this challenge, we propose using a data-driven networks approach that flags groups of corporate insiders who report coordinated transactions that are indicative of insider trading. Specifically, we leverage data on 2.9 million trades reported to the U.S. Securities and Exchange Commission (SEC) by company insiders (C-suite executives, board members and major shareholders) between 2014 and 2024. Our proposed algorithm constructs weighted edges between insiders based on the temporal similarity of their trades over the 10-year timeframe. Within this network we then uncover trends that indicate insider trading by focusing on central nodes and anomalous subgraphs. To highlight the validity of our approach we evaluate our findings with reference to two null models, generated by running our algorithm on synthetic empirically calibrated and shuffled datasets. The results indicate that our approach can be used to detect pairs or clusters of insiders whose behaviour suggests insider trading and/or market manipulation.}




\maketitle

\section{Introduction}\label{sec1}

The most common form of insider trading occurs when a corporate insider (i.e., an employee, director, or officer) exploits their role in an organisation to access and profitably trade on information that is not available to the general public. It is estimated that this type of insider trading, known as classical insider trading, accompanies 1-in-5 mergers \& acquisitions and 1-in-20 earnings announcements (source). Despite its prevalence, it remains notoriously difficult to detect, with the U.S. Securities and Exchange Commission (SEC) prosecuting merely 34 cases in 2024 \citep{jones2024insidertrading}. Considerable research has been dedicated to its adverse effects, with studies linking it to reduced liquidity in financial markets and undesirable impacts on managerial incentives \citep{fishman1992insider, bhattacharya2014insider}. As such, detecting and curbing insider trading is of clear interest to both governments and the general public.

To disincentivize the practice and increase transparency in financial markets, the US, alongside most other countries, legally requires that high-ranking corporate insiders report purchasing or selling any financial security issued by the companies they are affiliated with to the SEC within two working days of the transaction through a Form 4 filing \citep{sec_insider_forms}. This includes C-suite executives (e.g., CEOs and CFOs), members of the board of directors, and major shareholders. It should be noted that these trades are generally not illegal; company insiders tend to receive a substantial part of their compensation in the form of stock and stock options. Within specific windows, they are allowed to sell these securities or purchase additional ones; a transaction becomes illegal when an insider trades on material non-public information that their position gives them access to \citep{fried2013insider}.

Regulators use a number of tools to uncover insider trading, namely cross-market surveillance systems (CAT), targeted data requests (Electronic Blue Sheets), and tips/referrals, reported through the SEC’s whistleblower programme \citep{arif2022audit}. Finally, Form 4 disclosures have been used to screen for clustered trades around market-moving events, late filings, and coordinated patterns across related insiders \citep{tamersoy2014large}. Standard practice is event-based anomaly detection, where investigators screen for abnormally profitable trades and unusual spikes in volume in short windows around market-moving events \citep{priyadarshi2024comprehensive}. Accounts flagged in these windows are then prioritised for follow-ups by investigators.

The primary limitation of this approach is its focus on identifying individuals. This makes effective prosecution difficult since authorities must prove access to and use of material non-public information to meet the burden of showing scienter, while defendants can plausibly attribute gains to skill, luck, or public information. To address this gap, we propose a network-based screening method that flags pairs and clusters of corporate insiders who exhibit consistent and statistically unlikely temporal coordination in their trading patterns. From a prosecutor's perspective, the approach provides complementary evidence to existing methods, enabling group-level prosecution, where joint liability and cross-corroboration raise the likelihood that at least one insider cooperates.

We leveraged data on 2.9 million trades reported to the SEC by company insiders in the US between 2014 and 2024 to derive a network where edge weights between corporate insiders represent the temporal similarity of their trading patterns \citep{sec_insider_forms}.  Our method is built on the premise that if two or more insiders are consistently reporting sales and purchases within the same timeframe, then they could either be exchanging information or trading with reference to insider information that both parties have access to. The distinction here is that the former involves some form of communication, while the latter simply requires that insiders with access to the same information use it to make similar trades \citep{cohen2012decoding}. Links in this network can only be generated between two individuals who are affiliated with the same company at a given point in time, invariably giving rise to a relatively fragmented network. Nevertheless, connected components can consist of nodes from distinct companies if an insider occupies seats on multiple boards or if they worked for multiple companies during the 10-year period, both of which are important drivers of information dissemination in corporate networks \citep{ahern2017information}.

Our algorithm is (i) explainable, (ii) scalable, and (iii) generalisable in that it can be applied to derive a graph-based representation of coordination for any sequence of time-stamped trades. To assess its validity, we evaluate the resulting network against two generative null models. The first is a synthetic but empirically calibrated dataset and the second is a shuffled version of the actual data. Aspects of our empirically observed network that remain significant against both null models therefore represent coordination that is attributable neither to regulatory compliance nor coincidental responses to exogenous signals. We then inspected the actual network for suspicious activity with reference to central nodes and anomalous egonets, which reveals interesting findings, including two clusters of family members who consistently coordinate their trades. By demonstrating how data-driven networks capture high-order relationships over extended timeframes, our work offers an innovative framework for anti-financial crime investigations. This method yields interpretable, scalable insights into anomalous behaviour without requiring labelled training data.

\section{Related Work}

Network analysis offers a natural and effective framework for representing transactional data. Many scholars have therefore emphasised its value in strengthening the operational capacity of financial intelligence units, particularly in efforts against money laundering, terrorist financing, and fraud detection \citep{basu2018novel, li2022internet, oliveira2025complex}. Correspondingly, a growing body of research has been dedicated to studying networks of corporate insiders. This discourse is largely confined to two areas of research. One area focuses on how insiders' connections within the more general corporate environment correlate with the profitability of their trades \citep{el2015ceo, goergen2019insider}. These works ask whether well connected board members with access to information from multiple companies in the same industry report systematically more profitable trades than executives with less corporate clout. The second area regards how a corporate insider’s role within the company impacts the stock price upon the disclosure of a trade \citep{berkman2020inside, afzali2021network}. These studies build on the premise that the public disclosure of trades by high-ranking corporate executives can have a material impact on the stock price \citep{el2021network}. They then focus on evaluating whether trades reported by certain insiders have a more significant effect (e.g., trades by the CTO vs. the CEO).

To our knowledge, only two studies have previously leveraged graph-based data mining techniques to identify trends of insider trading within this data  \citep{tamersoy2014large, kulkarni2017network}. Both of these works use Form 4 filings to construct weighted edges between individuals based on how often they trade on the \textit{same day}. Our work follows the same rationale, but we address various limitations in their approach. First, instead of requiring exact matches, we apply a weekly linear kernel that gives partial credit to trades a few days apart, making the measure robust to reporting frictions and small timing discrepancies. Second, rather than counting exact date overlaps on sets, we use a match-based similarity function that aligns each trading day to its closest counterpart using a bounded weekly kernel, thereby avoiding combinatorial inflation from repeated near-coincidences and returning a continuous score in 
\([0,1]\). Third, we compute purchase- and sale-based similarities separately and combine them using activity-based weights, so that alignment in either contributes to the overall edge weight. Finally, recognising that insider trading windows span 120-160 days per year depending on the company, we adopt strict edge-creation criteria to curb coincidental co-occurrences \citep{guay2021insider}.

Further, since these previous attempts do not go beyond applying their algorithms to the empirically observed data, they cannot validate the statistical significance or reliability of their approaches. This limitation is critical in a setting where ground-truth labels for illicit activity are essentially non-existent, rendering standard supervised evaluation metrics inapplicable \citep{yeung2025garbage}. To address this, we move beyond descriptive analysis by benchmarking our findings against two complementary null models. First, we use a calibrated generative model that controls for structural constraints such as trading windows, insider tenure, and firm lifespans. This allows us to assess whether coordination can be explained purely by regulatory timing. Second, we use constrained temporal shuffling of the underlying data, which preserves the exact sequence of firm-level activity but randomises insider identities, allowing for an evaluation of whether synchronisation merely reflects common responses to exogenous factors other than insider trading. Within the context of the existing literature, our contributions can therefore be summarised as follows: 

\begin{itemize}
    \item We introduce an interpretable, scalable, and generalisable network-based screening method that quantifies coordinated trading behaviour among corporate insiders with reference to time-stamped Form~4 disclosures.
    \item We validate the approach using two complementary null models, demonstrating that the network’s structure reflects genuine coordination rather than coincidental overlap. By controlling for both systemic constraints and random coincidence, this method provides a statistical basis for identifying suspicious activity in noisy, unlabelled datasets more generally.
    \item We apply the method to 2.9~million US insider trades (2014--2024) and uncover previously unreported clusters of coordinated insiders, demonstrating its practical value as a forensic tool.
\end{itemize}

\section{Method}\label{sec3}

\subsection{Data}\label{sec3}

US federal law requires that corporate insiders report all changes in their beneficial ownership of securities to the SEC within two business days of the trade via a Form 4 filing. This requirement applies specifically to directors, C-suite executives, and shareholders owning more than 10\% of equity in a publicly traded company \citep{sec_insider_forms}. To analyse stock purchases and sales we focused on part 1 of this form, which is publicly available through the SEC's Electronic Data Gathering and Retrieval (EDGAR) system. Data on roughly 2.9 million trades reported between 2014 and 2024 was used to conduct the analysis. Trades reported by institutional investors (including Limited Liability Companies, Limited Partnerships, and Incorporated Companies) were removed to maintain a focus on individuals without having high-frequency transactions and large volumes distort the analysis. Additionally, we aggregated multiple line items that refer to the same insider, issuer, direction, and calendar day into single events. Summary statistics for the data are provided in Table~\ref{tab:insider_summary}.

\setlength{\textfloatsep}{5pt}

\begin{table}[!htbp]
\begin{tabular*}{\linewidth}{@{}l@{\extracolsep{\fill}}r r@{}}
\toprule
\textbf{Statistic} & \textbf{All trades} & \textbf{Aggregated by day} \\
\midrule
Number of insiders & 70,941 & — \\
Number of companies & 9,426 & — \\
Number of trades & 2,735,932 & 461,785 \\
Purchases & 1,326,228 & 243,640 \\
Sales & 1,409,704 & 218,145 \\
Average trades per insider & 38.6 & 6.51 \\
\botrule
\end{tabular*}

\caption{\textbf{Summary statistics for the raw insider trading data (2014–2024).} Purchases and sales are of similar magnitude, and daily aggregation reduces the number of observations by roughly a factor of six.}
\label{tab:insider_summary}

\end{table}

\subsection{Deriving the network}\label{subsec-deriving}

\paragraph{Temporal alignment between individuals}

Our method aims at deriving edge weights that capture the similarity of company insiders' trading patterns over the 10-year period under consideration. We first generated an empty network $G = (V, E)$, where $V$ and $E$ denote the sets of the nodes and connections between the nodes respectively. After which nodes and edges are introduced by comparing the transaction dates for every possible pair of insiders from the 9,426 firms. Let $X_C=\{x_1,\dots,x_m\}$ and $Y_C=\{y_1,\dots,y_n\}$ denote trade dates for two insiders at company $C$. Since we are defining edges between distinct nodes, we strictly require $X_C \neq Y_C$. We measured temporal proximity via a nonnegative weight $w(d)$ that decreases with the absolute time gap $d=|x-y|$.
Specifically, we used a linear ``same-week'' kernel

\begin{equation}
w(d)=
\begin{cases}
1-\dfrac{d}{7}, & d\le 7\\[4pt]
0, & d>7
\end{cases}.
\label{eq:weekly-kernel}
\end{equation}
In this way, events that are more than one week apart do not contribute to the similarity score, while a one--day gap receives weight $6/7$, and a same--day match receives weight $1$. In our context, linear decay offers a transparent, interpretable way to reduce similarity for temporally distant trades while maintaining a focus on near-simultaneous activity. With the kernel defined in \eqref{eq:weekly-kernel}, we can now compute the alignment between insiders. For each $x\in X_C$ we match to the most aligned event in $Y_C$, i.e.

\begin{equation}
s_{X\mid Y}
=\frac{1}{|X_C|}
\sum_{x\in X_C} \max_{y\in Y_C} w(|x-y|).
\end{equation}
This is the average, over $X_C$, of the best available proximity in $Y_C$. Similarly,
\begin{equation}
s_{Y\mid X}
=\frac{1}{|Y_C|}
\sum_{y\in Y_C} \max_{x\in X_C} w(|x-y|).
\end{equation}
Because $s_{X\mid Y}\neq s_{Y\mid X}$ if one insider trades more often, we define a symmetric measure by averaging both directions. For any pair of distinct insiders (where $X_C \neq Y_C$):
\begin{equation}
S(X_C,Y_C)
=\frac{1}{2}\!\left[
\frac{1}{|X_C|}\sum_{x\in X_C}\max_{y\in Y_C} w(|x-y|)
+\frac{1}{|Y_C|}\sum_{y\in Y_C}\max_{x\in X_C} w(|x-y|)
\right].
\label{eq:best-match}
\end{equation}

\noindent We computed \eqref{eq:best-match} separately for purchases $(A)$ and sales $(D)$:
\begin{equation}
S^{(A)} = S\!\big(X_C^{(A)}, Y_C^{(A)}\big), 
\qquad
S^{(D)} = S\!\big(X_C^{(D)}, Y_C^{(D)}\big).
\label{eq:purchases-sales}
\end{equation}
We then combined them with activity weighting,
\begin{equation}
S_{\text{combined}}
=\frac{|X_C^{(A)}|+|Y_C^{(A)}|}{T}\,S^{(A)}
+\frac{|X_C^{(D)}|+|Y_C^{(D)}|}{T}\,S^{(D)},
\label{eq:activity-weighted}
\end{equation}
where
\begin{equation}
T=|X_C^{(A)}|+|Y_C^{(A)}|+|X_C^{(D)}|+|Y_C^{(D)}|,
\label{eq:activity-weighted-denom}
\end{equation}
denoting the total number of trades made by the two insiders.

\paragraph{Interpretation and properties}
The final score \(S_{\text{combined}}\in[0,1]\) is a convex combination of \(S^{(A)}\) and \(S^{(D)}\), representing the expected weekly--kernel match rate of a random trade from the pooled logs. $S_{\text{combined}} = 1$ if every trade has a same--day counterpart, and $S_{\text{combined}} = 0$ if no matches occur within a week. The measure is symmetric, order--invariant, and reduces to the active component if the other is absent. This means that unmatched filings reduce the score, while matched same--week filings increase it. Tightening the weekly kernel lowers similarity, and although bounded and symmetric, the measure is a similarity rather than a metric. In the supplementary material we present an alternative approach to edge detection, which enforces one–to–one matching between trades. This stricter formulation avoids “burst inflation,” where multiple trades in one log align with the same counterpart, and provides a robustness check showing that our findings are not an artefact of the more permissive best–match scheme.

\begin{algorithm}[H]
\caption{Construct Insider Similarity Network}
\begin{algorithmic}[1]

\State \textbf{Create} empty graph $G = (V, E)$

\For{\textbf{each} company $C$}
    \For{\textbf{each} insider pair $(u,v)$ at $C$ where $u \neq v$}
        \State \textbf{Get} trade dates $Buys_u, Sells_u$ for insider $u$
        \State \textbf{Get} trade dates $Buys_v, Sells_v$ for insider $v$

        \State \textbf{Compute} $Sim_{Buy}$ using $\textsc{BestMatch}(Buys_u, Buys_v)$
        \State \textbf{Compute} $Sim_{Sell}$ using $\textsc{BestMatch}(Sells_u, Sells_v)$

        \State \textit{// Combine scores based on trade volume}
        \State \textbf{Set} $Score$ to the weighted average of $Sim_{Buy}$ and $Sim_{Sell}$

        \If{($u$ is active) $\land$ ($v$ is active) $\land$ 
            ($u$ has $\ge 8$ trades) $\land$ ($v$ has $\ge 8$ trades) $\land$
            (overlap $\ge 4$ weeks) $\land$ ($Score > 0.8$)}
            \State \textbf{Add} edge $(u,v)$ to $G$ with weight $Score$
        \EndIf
    \EndFor
\EndFor

\State \textbf{Return} $G$

\end{algorithmic}
\end{algorithm}

The similarity measure compares the complete set of reported trade dates for each pair of insiders belonging to the same company at a given point in time. We recognise that some overlap is expected to occur by chance since the windows within which company insiders are allowed to trade span roughly 120-160 days per year \citep{guay2021insider}. This means that it is to be expected that a given set of company insiders trade within the same weekly window at some point by chance. Accordingly, Equation~\eqref{eq:best-match} constructs an edge only if the following conditions are satisfied: (i) they have reported at least 8 trades; (ii) they have reported overlapping trades in at least 4 distinct weeks; and (iii) the similarity score \(S(X_C,Y_C)\) is greater than \(0.8\). Our supplementary material includes a proof that justifies this threshold by demonstrating that it requires a level of overlap that is statistically unlikely to happen by chance.

\subsection{Null models}

Robust null modelling can aid in measuring the likelihood that coordinated trading occurs purely by chance. To this end, we leveraged two distinct methods for generating synthetic versions of the underlying data to which we then applied our algorithm. First, we implemented a \textit{calibrated generative model} to control for regulatory artifacts. By uniformly distributing synthetic trades within admissible open windows (accounting for blackout periods and tenure), this model establishes a baseline for edge weights arising solely from shared trading constraints \citep{gauvin2022randomized}. Second, to disentangle specific social coordination from environmental homophily (e.g., collective reactions to public firm news), we utilised \textit{constrained temporal shuffling}. By randomly permuting insider identities within fixed firm-time bins while preserving exact transaction timestamps, this approach isolates dyadic synchronisation that exceeds the background level of concurrent firm activity.

\subsubsection{Calibrated generative model}

The first null model constitutes a \textit{randomised reference model}, constrained by empirical activity spans to determine whether observed temporal patterns are more or less likely than in a “random but comparable” world \citep{gauvin2022randomized}. This approach is standard practice in forensic finance for distinguishing skill (or information) from regulatory coincidence by generating a distribution of expected ``random co-occurrences'' that can be compared to the empirically observed values \citep{cohen2012decoding, fagiolo2013null}. For our purposes, the calibrated null model simulates a legally plausible but uncoordinated world of insider trading where (a) each insider trades as frequently and over as long a period as in reality, and (b) each firm retains the same total activity and lifetime as observed in the empirical data.

A table containing the full list of calibrated parameters can be found in the supplementary material. Conceptually, firm lifespans and insider tenures are anchored to their real first and last trade dates on the 2014-2024 timeline, which should be seen as a conservative proxy since insiders may have been legally eligible to trade before and after these observed bounds. Within these anchored spans, every insider–firm pair maintains its real number of trades and tenure duration, while the global proportion of purchases and sales is also preserved. Trades are redistributed uniformly across legally plausible open trading days, defined as the first 30 business days of 91-day quarters. We recognise that since legal trading windows are chosen by firms, they vary in length from roughly 25 to 40 days; we chose 30 days with reference to research from \cite{guay2021insider}, who reported an average of 30-35 trading days for US companies. Next, all quarters are assumed to begin on the same calendar dates for all firms. This does not impact co-trading within individual firms, but it slightly increases the likelihood of coincidental overlap between directors who sit on multiple boards. The model thus represents a legally consistent, empirically grounded, but uncoordinated world of corporate trading, thereby providing an appropriate baseline for asking: Are insiders’ real trading patterns, in terms of temporal overlap, more structured than they would be under random independent activity?

\subsubsection{Constrained temporal shuffling}

The synthetic randomised reference model establishes a useful baseline for evaluating the statistical significance of observed coordination in our network. Nevertheless, it ignores the fact that financial activity is often driven by exogenous public signals (e.g., current events, economic policy, and societal trends) that affect all insiders at a firm simultaneously \citep{cline2017persistence}. A cluster of trades may therefore appear coordinated not as a result of collusion, but simply because multiple insiders independently reacted to the same public event in a similar manner. To account for such non-stationarity, we leveraged a constrained randomised reference model, also referred to as a block-constrained configuration model \citep{gauvin2013activity, casiraghi2019block}. Methodologically, this stratified shuffle of the underlying data creates a counterfactual history by holding the precise timeline of all trading events constant while randomly permuting the identities linked to trading timestamps within each firm at the quarterly level.

Formally, we let the set of all trades be denoted by $\mathcal{E} = \{(u_i, f_i, t_i, d_i)\}$, where $u$ is the insider, $f$ is the firm, $t$ is the exact timestamp, and $d$ is the trade direction. We partitioned $\mathcal{E}$ into disjoint subsets $S_{f, \tau}$ based on the firm $f$ and a coarse-grained time-bin $\tau$ (e.g., calendar month). Within each subset $S_{f, \tau}$, we define the vector of insider identities $\mathbf{U}_{f,\tau} = [u_1, u_2, \dots, u_k]$. We generated the null dataset by applying a random permutation $\sigma$ to $\mathbf{U}_{f,\tau}$, such that the $j$-th trade in the null model is assigned to insider $u_{\sigma(j)}$, while the timestamp $t_j$ and direction $d_j$ remain fixed. This model preserves the exact temporal volume profile of the firm (including all bursts of activity potentially caused by exogenous factors) but destroys specific actor-to-actor synchronisation.  The approach takes inspiration from \cite{crane2019institutional}, who used a null model that randomised the links between investors while controlling for their industry and size preferences to better understand information sharing.

\subsection{Network analysis}

Upon deriving our observed network and assessing its structure with reference to the two above outlined null models, we inspected groups of people within it using closeness centrality, eigenvector centrality, and the OddBall algorithm, which was developed by \cite{akoglu2010oddball} to detect anomalous egonets (i.e., subgraphs formed by a node’s immediate neighbours) in weighted graphs. We started by ranking all nodes by their closeness centrality to capture insiders who are most 'suspicious' at a basic level. Following \cite{goergen2019insider}, we also rank nodes eigenvector centrality to identify insiders who are well embedded within the broader network by capturing not only how many nodes are linked to the target node but also their importance in the broader network (i.e., the degree of each connected node).

Finally, to identify anomalous nodes in graphs with weighted edges -- as is the case in our context -- \cite{akoglu2010oddball} first outlined four empirical patterns or power-law relationships that they generally see as innate to real-world weighted graphs. They then proposed an algorithm capable of identifying egos (nodes) whose egonets deviate from this structure. The method is suitable in our context since we are interested in identifying abnormal pairs or clusters of insiders within the broader structure. To confirm the validity of the approach, the supplementary material contains statistical tests and visualisations that prove all four conditions are satisfied by our data. With reference to notation from \cite{akoglu2010oddball}, the out-line score is given by:

\begin{equation}
\textit{out-line}(i) = \frac{\max(y_i, Cx_i^\theta)}{\min(y_i, Cx_i^\theta)} \cdot \log\left(|y_i - Cx_i^\theta| + 1\right),
\label{eq:outline}
\end{equation}

\noindent where $y_u$ is the number of edges in the egonet of ego $u$, and $f(x_u)$ is the expected number of edges according to the power-law fit, when egonet $u$ has $x_u$ nodes. The equation essentially ranks all nodes by their distance from the expected fit. Following \cite{akoglu2010oddball}, the final outlier score for $u$ is then derived by combining its out-line score (as defined above) with the Local Outlier Factor (LOF), a density-based measure that flags outliers in relatively sparse regions of the graph. This helps avoid false negatives for egos that are distant from the majority of points, but still close to the power-law trend. Outlier scores for all nodes are then ranked, after which high-ranking (i.e., abnormal egonets) can be assessed separately in greater depth.

\section{Results}\label{sec2}

\subsection{Network parameters}

The network, derived with reference to the similarity function presented in Section~\ref{subsec-deriving}, consists of 4,650 nodes and 7,007 edges. We only included nodes with at least one edge and since most company insiders reported trades that had little to no overlap with other insiders, they will not have any edges. The result is that we see significantly fewer nodes than people in the original data.

\begin{figure}[H]
  \centering
  \includegraphics[width=1.0\textwidth]{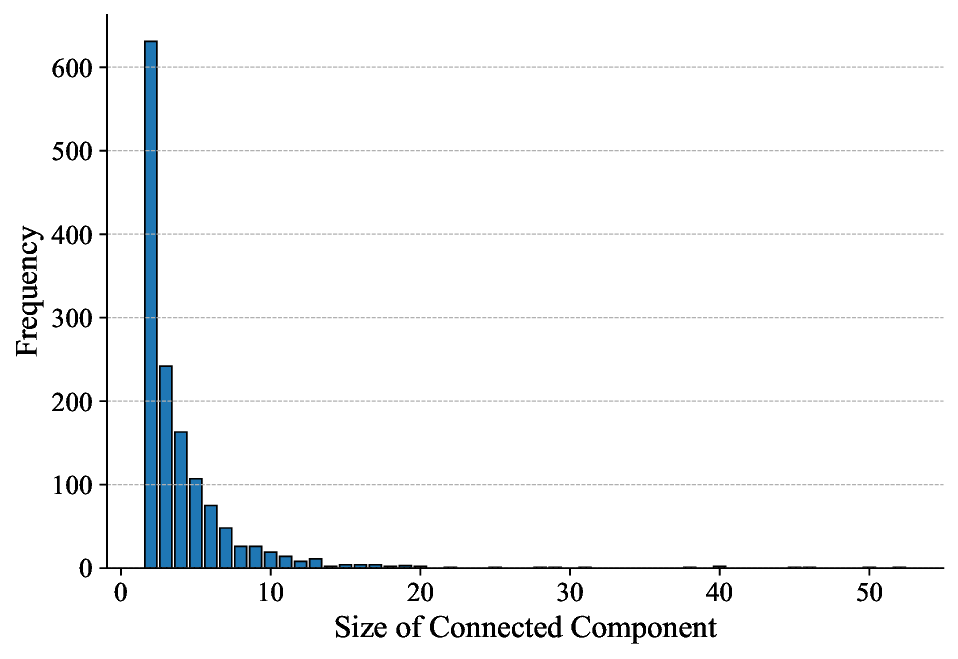}
  \caption{Frequency distribution of connected components by size in the network.}
  \label{fig:connected_components}
\end{figure}

Figure~\ref{fig:connected_components} reveals that the network is very sparse, with most of the 1,313 connected components consisting of only two nodes, corresponding to a density of 0.000648. This indicates that the majority of company insiders are not trading within similar timeframes, or at least not sufficiently similar to warrant an edge under our proposed similarity function. Nevertheless, Figure~\ref{fig:connected_components} also reveals some large components, including 10 clusters that comprise more than 30 corporate insiders. A table containing the distribution of connected components by the number of companies included can be found in the supplementary material. In this regard, it is worth noting that 93\% of all components consist exclusively of individuals associated with one firm. The remaining 7\% of connected components contain insiders from up to nine distinct companies.

\subsection{Null models}

Without establishing a baseline for random or coincidental activity, it is impossible to determine whether the above described patterns represent genuine collusion or merely reflect regulatory constraints and/or shared responses to exogenous events. To this end, our null framework benchmarks the network topology expected under the hypothesis of independent execution. For both null models, we generate $10^3$ synthetic datasets. We then apply our similarity function to each synthetic dataset and compare their structural properties with those of the empirical network.

\begin{table}[ht]
\centering
\small
\setlength{\tabcolsep}{3pt}
\begin{tabular}{l c c c c c}
\toprule
\textbf{Network Metric} & \textbf{Obs.} & \textbf{Structural Null} & \textbf{Z-Score} & \textbf{Shuffled Null} & \textbf{Z-Score} \\
& ($X_{obs}$) & (Mean $[5\%, 95\%]$) & ($Z_{struc}$) & (Mean $[5\%, 95\%]$) & ($Z_{shuff}$) \\
\midrule
\multicolumn{6}{l}{\textit{Global Connectivity}} \\
\hspace{2mm}Nodes & 4,650 & 21.9 $[15.0, 29.0]$ & 1,075$^{***}$ & 779.6 $[744, 813]$ & 178.6$^{***}$ \\
\hspace{2mm}Edges & 7,007 & 19.1 $[14.0, 24.0]$ & 2,271$^{***}$ & 1,449.6 $[1,401, 1,499]$ & 186.1$^{***}$ \\
\hspace{2mm}Average degree & 3.01 & 1.76 $[1.48, 2.13]$ & 6.1$^{***}$ & 3.72 $[3.57, 3.88]$ & -7.3$^{***}$ \\
\midrule
\multicolumn{6}{l}{\textit{Network Structure}} \\
\hspace{2mm}Connected components & 1,313 & 8.99 $[6.0, 12.0]$ & 651.8$^{***}$ & 258.9 $[244, 274]$ & 112.2$^{***}$ \\
\hspace{2mm}Giant component & 40 & 4.42 $[3.0, 6.0]$ & 33.7$^{***}$ & 41.0 $[41.0, 41.0]$ & --$^{\dagger}$ \\
\hspace{2mm}Multi-firm insiders & 23 & 0.59 $[0.0, 2.0]$ & 23.8$^{***}$ & 3.95 $[3.0, 7.0]$ & 15.4$^{***}$ \\
\bottomrule
\end{tabular}
\vspace{2mm}

\caption{\textbf{Comparative analysis of network topology against calibrated and shuffled null models.} The table reports empirical network metrics ($N=4,650$ nodes, $M=7,007$ edges) alongside the mean and 5--95\% confidence intervals derived from $10^3$ simulations of the calibrated null and the shuffled null. The Empirical Value ($X_{obs}$) is tested against the structural baseline ($H_{0}^{struc}$) to determine if topology arises from volume constraints, and against the shuffled baseline ($H_{0}^{shuff}$) to determine if it arises from homophily. $Z$-scores denote the standardised distance between the empirical value and the null distribution. Significance is indicated by asterisks ($^{***}\,P<0.001$) and for $^{\dagger}$ Null variance was zero, returning an undefined Z-score.}
\label{tab:network_topology}
\end{table}

Table \ref{tab:network_topology} quantifies the structural divergence of the observed insider network from the theoretical baselines. We employed $Z$-scores to standardise the magnitude of these deviations, allowing for direct comparison across metrics with vastly different scales. Furthermore, we report 95\% confidence intervals rather than standard deviations since null distributions in networks are often skewed \citep{squartini2011analytical}. As indicated in Table \ref{tab:network_topology}, the empirical network exhibits connectivity and cohesion that defy the counterfactual scenario. The observed number of active nodes (4,650) and edges (7,007) exceeds the structural baseline expectation by over 1,000 standard deviations ($Z_{struc} > 1,000$), confirming that the observed clustering is not a mechanical artifact of firm listing periods. Crucially, this significance is also inherent in the stricter `shuffled null' ($Z_{shuff} \approx 170$). This implies that at least some of the anomalous network structures we discuss in Section~\ref{subsec:empirical} are driven by specific, conscious coordination rather than aggregate market trends. Detailed descriptive statistics for the null distributions are provided in the supplementary material.

Although almost all metrics significantly exceed both null baselines, the largest component value for the shuffled model remained unchanged at a value of 40 across all shuffles. This number highlights the importance of using two null models in our context. Indeed, the calibrated null reveals that a connected component of size 40 is near impossible with the 95\% confidence interval bounded at 6. Conversely, in the shuffled null, the size of this component remains unchanged across all $10^3$ simulations, implying that all 40 insiders within this company are coordinating their behaviour, so shuffling them has zero influence on how our algorithm interprets their behaviour. This cluster will be discussed in greater depth in Section~\ref{subsec:empirical}. 

\newpage

\begin{figure}[ht]
  \centering
  \includegraphics[width=1.0\textwidth]{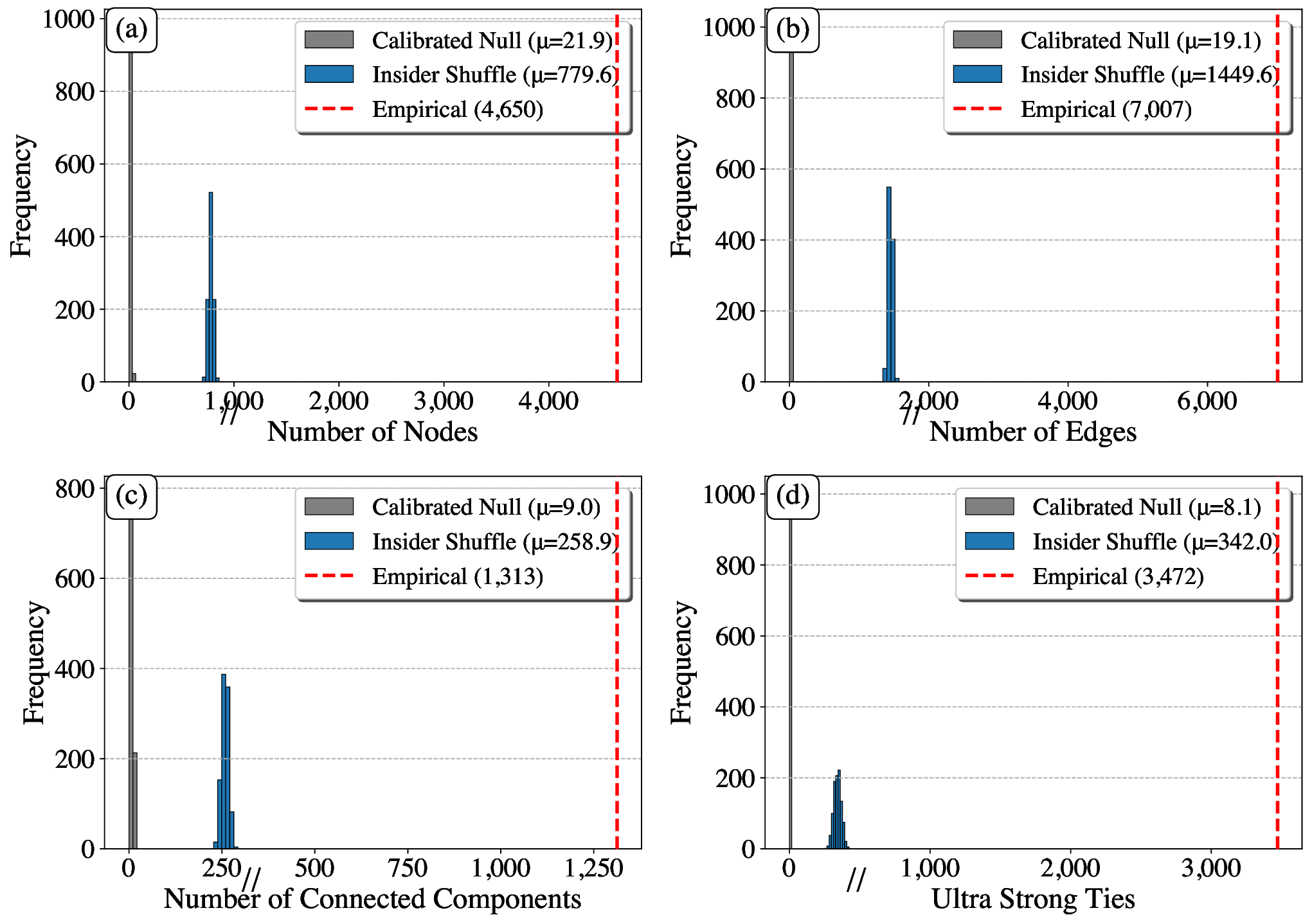}
  \caption{\textbf{Distribution of key network metrics from $10^3$ simulations of two null models compared to the empirical network.} Panels show (a) number of nodes, (b) number of edges, (c) number of connected components, and (d) ultra-strong ties (edges with similarity $> 0.9$). The $//$ on the x-axis denotes a discontinuity in the scale, used to visualise the tightly concentrated null distributions alongside the distant empirical values.}
  \label{fig:rich_club_metrics}
\end{figure}

Figure~\ref{fig:rich_club_metrics} visualises the distribution of four key topological metrics generated by $10^3$ simulations of the calibrated null (structural baseline, grey) and the insider shuffle (blue) against the empirically observed values (red dashed line). The results demonstrate a fundamental separation between the observed level of coordination and the counterfactual baselines. Indeed, the empirical values for all our metrics, which capture the scale of the active network (nodes, edges) and its cohesion (connected components, ultra-strong ties), lie far beyond the null distributions. For example, the observed network contains 3,472 ultra-strong ties ($w_{ij} > 0.9$), exceeding the shuffled expectation ($\mu \approx 340$) by an order of magnitude. This confirms that the high density of coordinated trading is not an artifact of monthly seasonality or sector-wide momentum, which are preserved by the shuffle.

\subsection{Empirical inspection}\label{subsec:empirical}

\subsubsection{Case study 1: Closeness centrality}

In this section, we conduct an interpretable and accessible inspection of the empirically derived network. We outline three ways of analysing the structure and use them to qualitatively discuss six of the most noteworthy egonets. The aim is to illustrate how financial authorities can use our algorithm to generate a network that can be used to identify groups of corporate insiders that warrant closer examination. We began with closeness centrality since nodes with the highest closeness scores are, by definition, those that can reach the rest of the network most efficiently.

\begin{figure}[ht]
\centering
\includegraphics[width=1\textwidth]{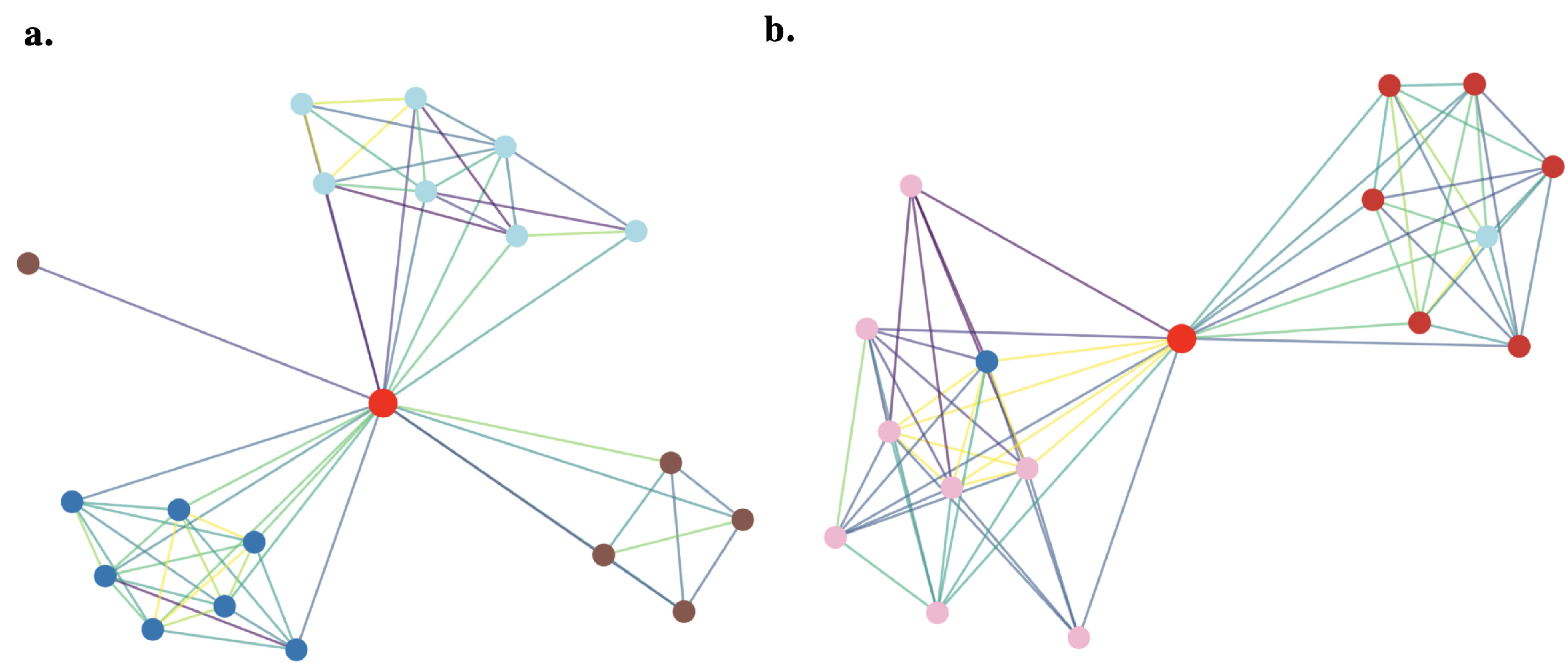}
\caption{Egonets of the two most central insiders measured via \textit{closeness centrality}. Node colours represent company affiliation, and edge colours capture edge weights, with yellow indicating strong ties, green medium-to-strong ties, blue moderate ties, and purple weak ties.}
\label{fig:closeness-central}
\end{figure}

Figure~\ref{fig:closeness-central} depicts the two insiders with the highest closeness centrality scores. Specifically, \hyperref[fig:closeness-central]{Panel~\ref*{fig:closeness-central}a} depicts an individual who occupies board seats at four mid-cap firms (market capitalisations between 4 and 12 billion USD) and reports substantial overlap with other executives across the four firms they are affiliated with. Next, \hyperref[fig:closeness-central]{Panel~\ref*{fig:closeness-central}b} shows an insider who served in various C-suite roles across five different companies throughout the 10-year period under consideration. For example, the large pink cluster to which they are strongly connected represents board members of a Fortune 500 company. It is also important to note that both individuals maintain ties to merely a subset of directors across their respective boards, with edge weights varying considerably. This variation suggests that the observed clusters cannot be attributed to insiders receiving equity compensation at the same point in time.

We cross-referenced our ranking of the 100 most central nodes against a comprehensive database of federal securities fraud prosecutions. This revealed that three insiders with exceptionally high centrality scores were board members at AVEO Pharmaceuticals, a company subject to high-profile fraud litigation regarding the concealment of feedback from the Food and Drug Administration. Crucially, the subsequent enforcement actions did not charge the specific board members included in our network, but rather the CEO, CFO, and CMO, who ultimately settled civil charges or were found liable for securities fraud \citep{sec2018aveo}. This means that although our algorithm did not pinpoint the exact perpetrators, it successfully identified the corporate network as a high-risk locus of informational asymmetry.

\subsubsection{Case study 2: Eigenvector centrality}

While closeness centrality captures individuals who can effectively reach the rest of the network, it does not account for the importance of neighbouring nodes. Eigenvector centrality addresses this limitation by assigning higher scores to insiders who are linked to well-connected peers. This makes it particularly useful for identifying structurally influential groups whose behaviour may propagate through the network or amplify coordinated trading patterns. In our case, the distribution of eigenvector centrality scores is highly skewed (see supplementary material): most insiders occupy peripheral positions, while a small subset exhibits markedly higher centrality.

\begin{figure}[H]
\centering
\includegraphics[width=1\textwidth]{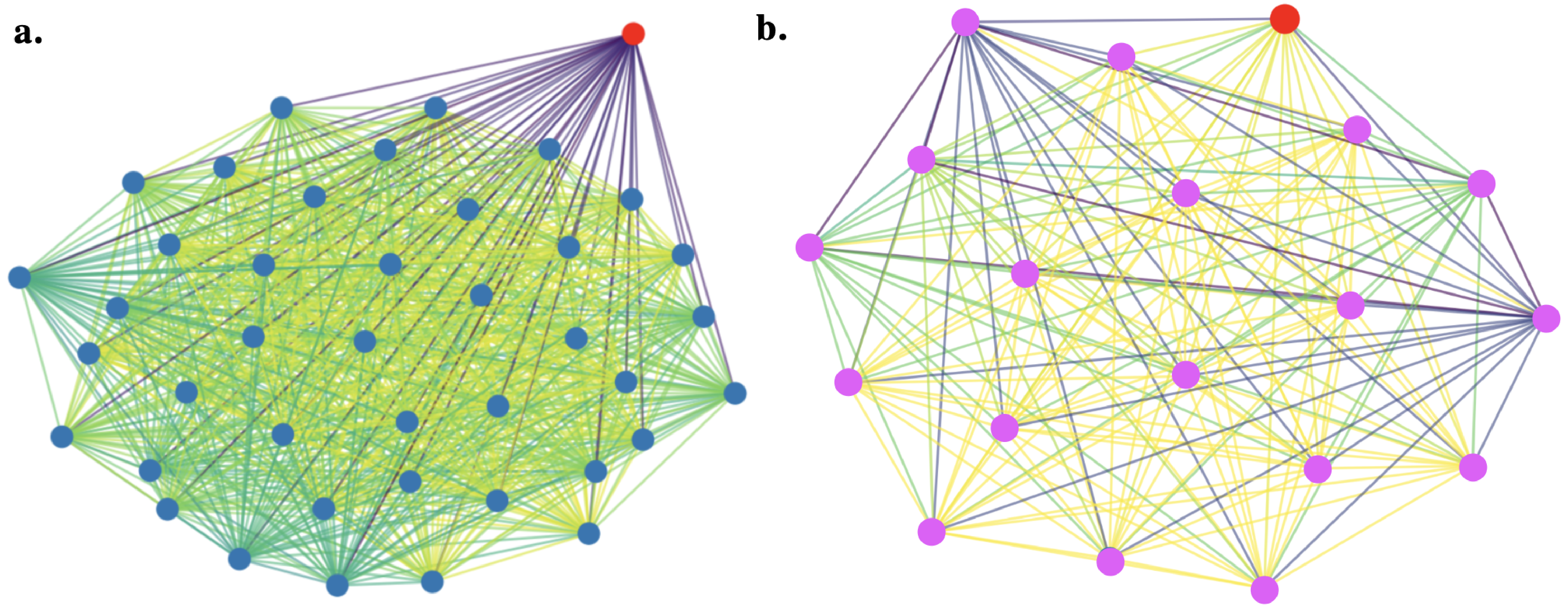}
\caption{Egonets of the two most central insiders measured via \textit{eigenvector centrality}. Node colours represent company affiliation, and edge colours capture edge weights, with yellow indicating strong ties, green medium-to-strong ties, blue moderate ties, and purple weak ties. The red node denotes the individual under consideration.}
\label{fig:eigenvector-central}
\end{figure}

 Figure~\ref{fig:eigenvector-central} depicts egonets for two nodes with the highest eigenvector centrality scores. The first connected component depicted here contains a total of 41 company insiders. Interestingly, 36 of these insiders are part of the same family, consistently trading large chunks of equity within similar timeframes. The family's synchronised trades raise  concerns regarding informational advantages, collusive behaviour, and limited oversight in corporate governance structures. With a market cap of less than USD 1 billion, transactions like this are also likely to have a material impact on the actual stock price \citep{gomez2024effect}. Interestingly, \hyperref[fig:eigenvector-central]{Panel~\ref*{fig:eigenvector-central}b} also depicts a cluster of insiders belonging to a large family who traded shares in a former motorsports entertainment company that was acquired by NASCAR in October 2019 \citep{isc2019merger}.

\subsubsection{Case study 3: Anomalous egonets}

Next, we examine anomalous egonets using the OddBall algorithm discussed earlier. OddBall identifies nodes whose local connectivity patterns deviate from the network’s expected structural relationships, allowing for the identification of groups whose trading behaviour appears atypical.

\begin{figure}[H]
\centering
\includegraphics[width=1\textwidth]{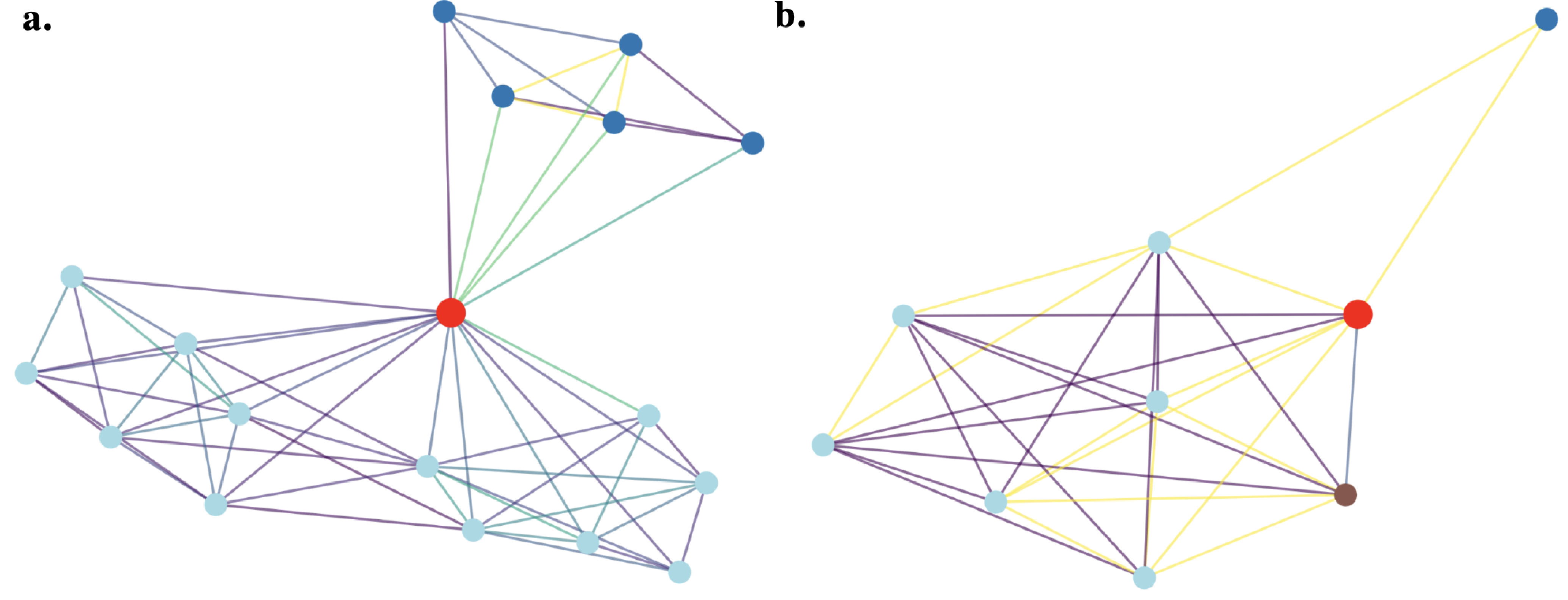}
\caption{Egonets of the four most anomalous individuals identified using the \textit{OddBall} algorithm. Node colours represent company affiliation, and edge colours capture edge weights, with yellow indicating strong ties, green medium-to-strong ties, blue moderate ties, and purple weak ties. The red node denotes the individual under consideration.}
\label{fig:anomalies}
\end{figure}

Remarkably, when cross-referencing the names of our 100 most anomalous egonets against data on past prosecutions, we found that the corporate insider with the most anomalous egonet has been prosecuted for insider trading in connection with a major accounting scandal. Specifically, the egonet in \hyperref[fig:anomalies]{Panel~\ref*{fig:anomalies}a} depicts the CEO of Plug Power Inc. who was a defendant in a consolidated class action lawsuit filed in the U.S. District Court for the Southern District of New York. The allegations paint a picture of classic financial engineering followed by opportunistic insider selling \citep{plug2022shareholderalert}. Importantly, the complaint alleges that shortly before these restatements were announced, the CEO and CFO sold approximately \$45 million of their personal holdings. Our algorithm picked up on this large liquidation event, generating a strong link between the two, which triggered the ``anomalous'' flag when running OddBall.

Next, \hyperref[fig:anomalies]{Panel~\ref*{fig:anomalies}b} depicts a prominent California-based venture capitalist who serves on multiple boards and reported overlapping trading activity with executives at three other firms. Notably, there is substantial overlap with four of the six board members to which this insider is connected to. A closer look at their transactions reveals that all four insiders repeatedly sold large blocks of stock (exceeding \$50,000) within two-day windows. This pattern may suggest trading on shared information. Alternatively, given the company’s market capitalisation of under \$500 million, the six insiders may be deliberately coordinating their sales within these windows before reporting them to the SEC, thereby avoiding the price impact that early disclosure of a single insider’s trade could trigger. This type of coordination is not implausible, given that financial markets are known to react to large trading disclosures from high-ranking corporate insiders \citep{fidrmuc2008insider, gomez2024effect}.

\section{Discussion}

The detection of insider trading has traditionally focused on identifying individual anomalies, with regulators and compliance officers screening for isolated accounts that generate abnormal returns \citep{goergen2019insider, priyadarshi2024comprehensive}. Our research demonstrates that this atomistic view misses a fundamental dimension of financial misconduct, namely the social and coordinated nature of information flow. We propose an algorithm that links insiders who repeatedly trade in the same short windows, using a weekly kernel and strict overlap thresholds to construct a robust similarity network. Its application to 2.9 million trades reported between 2014 and 2024 gives rise to a sparse but highly structured graph composed of numerous anomalous clusters containing tightly coordinated insiders.

The comparison to null models is central to the interpretation of our findings. They provide the baseline required to determine if the network captures genuine coordination or merely a shared response to market volatility in the absence of labelled data. First, the calibrated generative model preserves insider–firm tenures, firm lifespans, and overall trade counts while redistributing trades uniformly over admissible open days. This synthetic data represents a conservative baseline for ``legally plausible but uncoordinated'' insider activity. Under this model, we found that the probability of observing a network with comparable numbers of active nodes, edges, or ultra-strong ties is very small \citep{guay2021insider}. Second, the shuffled null model confirmed that these connections are not driven by collective reactions to exogenous factors. When we retain the exact volume profile of a firm but randomise the identities of the traders, the network structure collapses. This implies that at least some of the large clusters of insiders, especially those containing individuals from multiple companies, represent subversive synchronisation in trading behaviour.

At a more granular level, the empirical inspection of central and anomalous egonets illustrates the types of patterns that regulators and compliance teams are able to identify via the proposed algorithm. For example, high-eigenvector insiders were generally embedded in dense clusters, where multiple board members systematically traded within similar short windows. The family clusters we uncovered are particularly significant, with large groups of related insiders repeatedly executing sizeable coordinated trades, often in relatively small-cap firms where such trades are likely to have non-trivial price impact. The OddBall analysis reinforced these qualitative impressions, flagging egonets whose weighted structure deviates sharply from the power-law regularities observed in the broader network. In practice, these egos and their immediate neighbourhoods would constitute natural candidates for deeper forensic follow-up. Crucially, the fact that the most anomalous egonet identified by our method corresponds to an insider who has been prosecuted for insider trading provides a form ex post validation in a domain where labelled ground truth is largely unavailable. While our approach is not designed to predict individual culpability, this true positive indicates that the network-based signal captures behaviour that is materially aligned with regulatory notions of illicit coordination.

In the context of these findings, it is important to emphasise that our framework is designed as a screening and prioritisation tool, not as a standalone mechanism for establishing liability. Our null models help rule out purely statistical coincidence under plausible trading constraints, but they cannot distinguish between illegal exploitation of material non-public information and lawful joint responses to public information or internal governance decisions with certainty. A comprehensive implementation of our approach should therefore also incorporate data on stock prices to assess whether individuals `flagged' by the network are reporting abnormal returns. This would certainly be relevant given that \cite{cziraki2021dollar} observed that individuals typically refrain from insider trading unless the potential profits are substantial enough to justify the associated risks. As such, future work should incorporate a complementary analysis on the profitability of trades reported by central or anomalous insiders within the network. This would make for robust evidence in the prosecution of insider trading.

There are three noteworthy limitations to our work. First, our analysis is restricted to Form 4 filings, which means we only observe the ``visible'' layer of insider trading. The most egregious violators, who trade through offshore accounts or shell companies to avoid reporting requirements, are naturally absent from this dataset. Second, it disregards financial derivatives, including stock options and warrants. This is of relevance since high-ranking executives often receive large incentive packages in the form of stock options \citep{augustin2019informed}. Notably, since these also have to be reported by company insiders to the SEC, they could be incorporated into our equation or modelled as part of a separate network.

Third, our match-based similarity treats overlaps proportionally, such that a ratio of 9/10 is equivalent to 18/20. In practice, however, the absolute scale of overlap may carry information about the severity of coordination. Our three conditions for edge creation partly mitigate this issue by requiring minimum activity and multiple overlapping weeks but future efforts could incorporate severity scores that weight both relative and absolute overlap to better capture the strength of trading synchrony. Finally, from a methodological perspective, future works should also place greater emphasis on small but strongly connected components. Indeed, our analysis primarily focused on identifying clusters of insiders who are coordinating their trades. However, investigating connected components that merely contain two or three strongly connected nodes would also be of relevance to the objective, since these insiders are reporting heavily coordinated trades that deviate from others within the firm whose behaviour was not deemed sufficiently similar to warrant an edge. \cite{sun2024prediction} present some promising work in this regard by leveraging graph-based machine learning to identify patterns of insider trading in a network of S\&P 500 companies.

\section{Conclusion}

Our objective was to identify corporate insiders who have traded with reference to material non-public information that two or more parties had access to by detecting abnormalities in the temporal overlap of their trades. From a prosecutor’s perspective, this approach provides complementary evidence to traditional analyses of abnormal profitability. Conceptually, our work contributes to three strands of research. First, within the literature on the detection of insider trading, our approach complements traditional pre-announcement screens and profit-based anomaly detection by shifting attention from individual outliers to relational patterns of behaviour. Second, from a more general forensic network science perspective, our analysis demonstrates how carefully constructed null models can separate structural features attributable to institutional constraints (e.g., trading windows, tenure, firm lifespans) from those likely driven by coordinated decision-making. Third, methodologically, we showed that unsupervised, label-free anomaly detection can be rendered operational in highly noisy financial settings, where ground-truth labels for illicit behaviour are scarce or absent.

\newpage

\section*{Declarations}

\subsection*{Author contributions}
Gian Jaeger (G.J.), Wang Ngai Yeung (W.N.Y.) and Renaud Lambiotte (R.L.) are authors of this article. G.J conceived the study, conducted the experiments/simulations and wrote the first draft of the manuscript. R.L. and W.N.Y. supervised the project. All authors contributed to the revision and the final version of the manuscript.

\subsection*{Funding}
R.L. acknowledges support from the EPSRC Grants EP/V013068/1, EP/V03474X/1, and EP/Y028872/1.

\subsection*{Conflict of interest}
The authors declare that they have no competing interests.

\subsection*{Ethics approval and consent to participate}
Not applicable.

\subsection*{Data availability}
All data used in this study is publicly available from https://www.sec.gov/search-filings.

\subsection*{Material availability}
The article has accompanying supplementary material.

\subsection*{Code availability}
All code will be made available upon the article's official publication.





\newpage

\bibliography{sn-bibliography}

\newpage

\addtocontents{toc}{\protect\setcounter{tocdepth}{0}}
\appendix
\input{supplement.tex}

\end{document}

%% file: supplement.tex
\clearpage
\begingroup
\newgeometry{a4paper,margin=1in}

\renewcommand{\thepage}{S\arabic{page}}
\setcounter{page}{1}

\vspace{1em}
\begin{center}
\Large \textbf{Supplementary Material}
\end{center}

\vspace{-1em}


\section{Justification for the \texorpdfstring{$0.8$}{0.8} edge threshold}
\addcontentsline{supp}{section}{A\quad Justification for the $0.8$ edge threshold}

\subsection*{Set-up and notation}

We consider two insiders at the same company \(C\) with trade-date multisets \(X_C=\{x_1,\dots,x_m\}\) and \(Y_C=\{y_1,\dots,y_n\}\), where $X_C \neq Y_C$
Temporal proximity is measured by the \emph{weekly kernel}
\begin{equation}
w(d)=
\begin{cases}
1-\dfrac{d}{7}, & d\le 7,\\[4pt]
0, & d>7,
\end{cases}
\qquad d=|x-y|.
\label{eq:weekly-kernel-supp}
\end{equation}
Directional best–match similarities are
\[
s_{X\mid Y}
=\frac{1}{m}\sum_{x\in X_C}\max_{y\in Y_C} w(|x-y|),
\qquad
s_{Y\mid X}
=\frac{1}{n}\sum_{y\in Y_C}\max_{x\in X_C} w(|x-y|),
\]
and the symmetric best–match similarity is
\begin{equation}
S(X_C,Y_C)
=\frac{1}{2}\Bigl(s_{X\mid Y}+s_{Y\mid X}\Bigr).
\label{eq:best-match-supp}
\end{equation}

\paragraph{Edge pre-filters.}
Throughout, we impose the same edge filters as in the main text: each insider has at least \(8\) trades over the study period, and the pair exhibits overlaps in at least \(4\) distinct weeks. These filters strengthen the conclusions below but are not required for the probability bounds we derive.

\subsection*{Null model of chance overlap}

Let \(A\) denote the number of \emph{allowed} trading days over the 10-year horizon. We fix \(A=120\times 10=1200\) allowed days (see the remarks for a discussion of why this is conservative). Under the \emph{independent random timing} null, conditional on counts \((m,n)\), each insider chooses trade days independently and uniformly from the \(A\) allowed days (with replacement), independently across insiders.

\paragraph{A conservative reduction.}
For any fixed \(x\in X_C\),
\[
\max_{y\in Y_C} w(|x-y|)
\;\le\; \mathbf{1}\{\exists\, y\in Y_C:\ |x-y|\le 7\}.
\]
Thus \(m\,s_{X\mid Y}\) is stochastically dominated by a binomial variable that counts how many \(x\)-trades have \emph{some} weekly neighbour in \(Y_C\). Writing
\begin{equation}
p_n \;=\; \mathbb{P}\big(\exists\, y\in Y_C:\ |x-y|\le 7\big)
= 1-\Bigl(1-\frac{14}{A}\Bigr)^{n},
\label{eq:pn}
\end{equation}
we have
\begin{equation}
m\,s_{X\mid Y}\ \preceq\ \mathrm{Binomial}\!\big(m,p_n\big),
\qquad
n\,s_{Y\mid X}\ \preceq\ \mathrm{Binomial}\!\big(n,p_m\big).
\label{eq:binom-dom}
\end{equation}
This replacement is deliberately conservative: the indicator counts any within-week neighbour as full weight \(1\), whereas the linear kernel in~\eqref{eq:weekly-kernel-supp} down-weights gaps of \(1\)–\(7\) days, making high similarities strictly \emph{less} likely under chance.

\subsection*{Tail bounds and family-wise calibration}

For \(\tau\in(0,1)\) and \(K\sim\mathrm{Binomial}(k,p)\),
\begin{equation}
\mathbb{P}\!\left(\frac{K}{k}\ge \tau\right)
\ \le\ \exp\!\big\{-k\,D(\tau\Vert p)\big\},
\qquad
D(a\Vert b)=a\log\frac{a}{b}+(1-a)\log\frac{1-a}{1-b}.
\label{eq:chernoff}
\end{equation}
Since \(S=\tfrac12(s_{X\mid Y}+s_{Y\mid X})\ge \tau\) forces \emph{both} directions to be at least \(\tau\), \eqref{eq:binom-dom}–\eqref{eq:chernoff} yield
\begin{equation}
\mathbb{P}\big(S\ge \tau\big)
\ \le\ \exp\!\big\{-m\,D(\tau\Vert p_n)-n\,D(\tau\Vert p_m)\big\}.
\label{eq:S-tail}
\end{equation}

Let \(N_{\text{pairs}}=\sum_{C}\binom{n_C}{2}\) be the number of insider pairs within firms. With \(70{,}941\) insiders across \(9{,}426\) firms, the empirical average is \(\bar n\approx 7.53\) insiders per firm, giving
\[
N_{\text{pairs}}
\ \approx\ 9{,}426\times \binom{\bar n}{2}
\ \approx\ 2.31\times 10^5.
\]
We control the dataset-level family-wise error rate (FWER) at \(\alpha=0.05\) by requiring
\[
\mathbb{P}(S\ge \tau_\star)\ \le\ \frac{\alpha}{N_{\text{pairs}}}
\quad\text{for every eligible pair.}
\]
In the least-stringent case for our filters (\(m=n=8\)), \eqref{eq:pn} gives
\[
p_8 \;=\; 1-\Bigl(1-\frac{14}{1200}\Bigr)^8 \;=\; 0.08961.
\]
By \eqref{eq:S-tail}, it suffices to choose \(\tau_\star\) with
\[
D(\tau_\star\Vert p_8)\ \ge\ \frac{\log(N_{\text{pairs}}/\alpha)}{m+n}
\ =\ \frac{\log(2.31\times 10^5/0.05)}{16}
\ =\ 0.959\ \ (\text{to 3 s.f.}).
\]
Solving the binary KL inequality yields the family-wise calibrated critical similarity
\begin{equation}
\boxed{\ \tau_\star \approx 0.652\ \ \text{(FWER }\le 5\%\text{ under independent random timing).}\ }
\label{eq:tau-star}
\end{equation}

\subsection*{Remarks on conservatism}
\begin{enumerate}\setlength{\itemsep}{2pt}
\item \emph{Kernel conservatism.} Replacing the linear weekly kernel by a \(0\)–\(1\) window strictly increases similarity under the null, so all tail probabilities above are \emph{upper bounds}.
\item \emph{Activity filters.} The requirement of \(\ge 4\) distinct overlapping weeks is not used in the bounds and further suppresses spurious alignment from isolated bursts.
\item \emph{Larger activity counts.} For insiders with more than \(8\) trades each, the bound in \eqref{eq:S-tail} improves exponentially in \((m+n)\) via the KL term \(D(\tau\Vert p)\), making false positives even less likely.
\item \emph{Allowed-day horizon.} We fix \(A=1200\) (120 trading-eligible calendar days/year over 10 years), which is conservative: some firms afford \(120\text{--}180\) days/year, i.e., \(A\) up to \(1800\). Larger \(A\) lowers \(p_n\) in \eqref{eq:pn}, making chance overlaps rarer and reducing the calibrated \(\tau_\star\), thereby strengthening \eqref{eq:tau-star}.
\end{enumerate}

\subsection*{Interpretation}

Under a transparent and conservative null calibrated to the study horizon (10 years, \(A=1200\) allowed days) and our edge filters (each insider has \(\ge 8\) trades), the family-wise calibrated critical similarity is \(\tau_\star\approx 0.652\) as in \eqref{eq:tau-star}. Using \(S\ge 0.8\) as the edge threshold is therefore a non-arbitrary and highly conservative choice that renders chance edges effectively impossible at the dataset scale.

\newpage


\section{Alternative assignment-based equation for edge detection}
\addcontentsline{supp}{section}{B\quad Alternative assignment-based equation for edge detection}

Let $X_C=\{x_1,\dots,x_m\}$ and $Y_C=\{y_1,\dots,y_n\}$ denote the event times
(i.e., Form~4 transaction dates) for two insiders at company $C$
(purchases or sales). Temporal proximity is scored by a weekly kernel
\begin{equation}
w(d)=
\begin{cases}
1-\dfrac{d}{7}, & d\le 7,\\[6pt]
0, & d>7,
\end{cases}
\qquad d=|x-y|.
\label{eq:weekly-kernel2}
\end{equation}
Thus same--day events receive weight $1$, a one--day gap receives $6/7$, and
gaps greater than a week contribute zero.

\paragraph{One--to--one matching.}
We construct a maximum--weight bipartite matching
$M^\star\subseteq X_C\times Y_C$
with respect to the edge weights $w(|x-y|)$, subject to:
(i) each event is matched to at most one counterpart (injective constraint);
(ii) only pairs with $d\le 7$ are eligible; pairs with $d>7$ are excluded.
This one--to--one constraint prevents burst inflation by ensuring that a single
trade cannot simultaneously explain multiple trades by the other insider.

\paragraph{Window-level eligibility.}
Rather than normalising by the total number of events, we normalise by the
number of temporal windows in which coordination is possible.
Let $\mathcal W_X$ and $\mathcal W_Y$ denote the sets of calendar weeks in which
insiders $X$ and $Y$ trade, respectively, and define the set of
\emph{overlap--capable weeks}
\begin{equation}
\mathcal W_{XY} = \mathcal W_X \cap \mathcal W_Y.
\end{equation}
Weeks in which only one insider trades are treated as ineligible and do not
contribute to the denominator.

\paragraph{Directional scores.}
We define the $X$--given--$Y$ directional score as
\begin{equation}
s_{X\mid Y}
=\frac{1}{|\mathcal W_{XY}|}
\sum_{(x,y)\in M^\star} w(|x-y|),
\label{eq:dir-x-window}
\end{equation}
with the convention that $s_{X\mid Y}=0$ if $\mathcal W_{XY}=\varnothing$.
The $Y$--given--$X$ score is defined analogously,
\begin{equation}
s_{Y\mid X}
=\frac{1}{|\mathcal W_{XY}|}
\sum_{(x,y)\in M^\star} w(|x-y|).
\label{eq:dir-y-window}
\end{equation}

\noindent This normalisation implies that multiple trades within the same overlapping
week neither inflate nor dilute the score: once both insiders are active in a
given week, additional within--week activity is neutral.

\paragraph{Symmetric similarity (per category).}
We define the symmetric similarity measure as
\begin{equation}
S_{\mathrm{assign}}(X_C,Y_C)
=\tfrac{1}{2}\big(s_{X\mid Y}+s_{Y\mid X}\big),
\label{eq:assign-sim2}
\end{equation}
which satisfies $S_{\mathrm{assign}}\in[0,1]$ under
\eqref{eq:weekly-kernel2}.

\paragraph{Combining purchases and sales (activity--weighted).}
We compute \eqref{eq:assign-sim2} separately for sales ($A$) and
purchases ($D$):
\begin{equation}
S_{\mathrm{assign}}^{(A)}=S_{\mathrm{assign}}(X_A,Y_A),
\qquad
S_{\mathrm{assign}}^{(D)}=S_{\mathrm{assign}}(X_D,Y_D).
\label{eq:assign-cat2}
\end{equation}
Let
\[
T_A = |X_A|+|Y_A|,\qquad
T_D = |X_D|+|Y_D|,\qquad
T = T_A+T_D.
\]
We define the overall similarity as the activity--weighted average
\begin{equation}
S_{\mathrm{combined}}(X,Y)
= \frac{T_A}{T}\,S_{\mathrm{assign}}^{(A)}(X,Y)
+ \frac{T_D}{T}\,S_{\mathrm{assign}}^{(D)}(X,Y).
\label{eq:combined2}
\end{equation}

\paragraph{Interpretation and properties.}
\begin{itemize}
\item $S_{\mathrm{assign}}^{(\cdot)}\in[0,1]$ and
      $S_{\mathrm{combined}}\in[0,1]$.
\item $S_{\mathrm{assign}}^{(\cdot)}=1$ if, in every overlapping week, the
      insiders’ trades are perfectly aligned in time.
\item The one--to--one matching removes burst inflation: a single trade cannot
      generate multiple matches.
\item Normalisation by overlapping weeks ensures that dense within--week
      trading is neutral, while activity in non--overlapping weeks weakens the
      score.
\item The activity--weighted combination in \eqref{eq:combined2} reflects
      whether coordination occurs primarily in purchases or sales.
\end{itemize}

\section{Calibrated null parameters}\label{appendix:calibrated_params}
\addcontentsline{supp}{section}{C\quad Calibrated null parameters}

\begin{table}[h!]
\centering
\caption{Parameters of the calibrated null model}
\begin{tabular}{lll}
\toprule
\textbf{Parameter} & \textbf{Description} & \textbf{Value / Distribution} \\
\midrule
$N_f$ & Number of firms & 9{,}381 (empirical) \\
$N_i$ & Number of insiders & 65{,}001 (empirical) \\
$C_{if}$ & Trades per insider–firm pair & Empirical integer counts \\
$T_f$ & Firm activity span & Empirical first–last trade dates (anchored) \\
$T_{if}$ & Insider–firm tenure length & Empirical span in days \\
$p_{\text{buy}}$ & Probability of purchase ($A$) & 0.528 (empirical) \\
$\text{Horizon}$ & Simulation window & 2014--01--01 to 2024--12--31 \\
$Q$ & Quarter length & 91 calendar days \\
$D_{\text{open}}$ & Open business days per quarter & 30 \\
$W_{\text{open}}$ & Global open-day calendar & First 30 business days of each 91-day quarter \\
\bottomrule
\end{tabular}
\label{tab:null_parameters}
\end{table}

\section{Distribution of connected components by number of companies included}\label{appendix:connected_compo}
\addcontentsline{supp}{section}{D\quad Distribution of connected components by number of companies included}

\begin{table}[h]
\caption{Distribution of connected components by number of companies included.}\label{tab:component_distribution}
\begin{tabular*}{\linewidth}{@{}l@{\extracolsep{\fill}}rrrrrrrrr@{}}
\toprule
\textbf{Number of companies} & \textbf{1} & \textbf{2} & \textbf{3} & \textbf{4} & \textbf{5} & \textbf{6} & \textbf{7} & \textbf{8} & \textbf{9} \\
\midrule
Components (\%) & 93.3\% & 4.6\% & 0.9\% & 0.5\% & 0.3\% & 0.3\% & 0.1\% & 0.0\% & 0.1\% \\
\bottomrule
\end{tabular*}
\end{table}

\section{Distribution of eigenvectors}\label{appendix:eigenvectors}
\addcontentsline{supp}{section}{E\quad Distribution of eigenvectors}

\begin{figure}[H]
\centering
\includegraphics[width=0.8\textwidth]{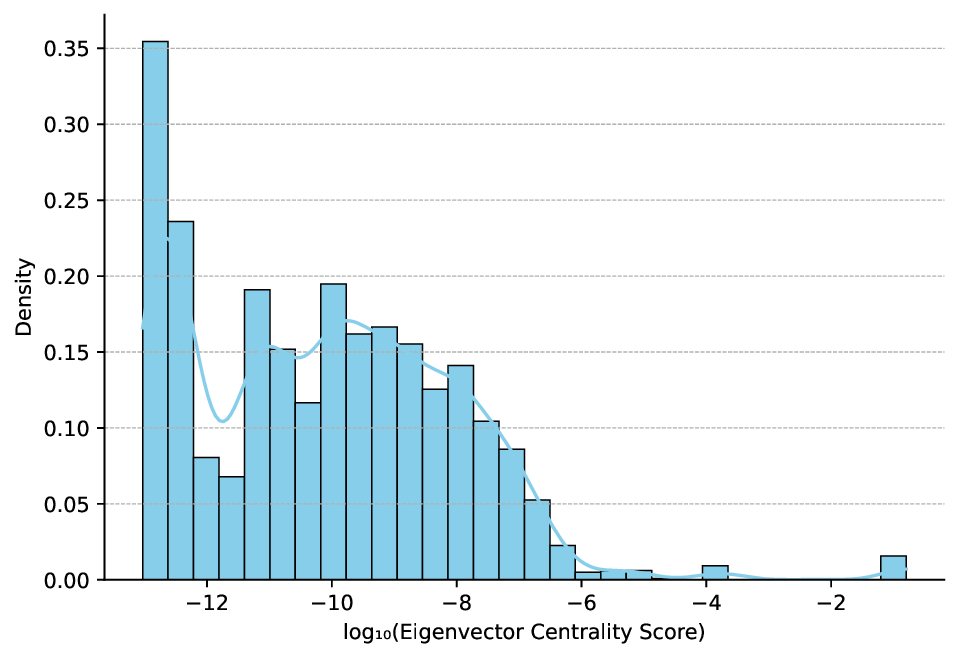}
\caption{Distribution of Eigenvectors (log scale)}
\label{fig:eigenvectors}
\end{figure}

\newpage

\section{Detailed topological statistics for null models}
\addcontentsline{supp}{section}{F\quad Detailed topological statistics for null models}

\begin{table}[h]
\centering
\setlength{\tabcolsep}{0pt} 
\begin{tabular*}{\textwidth}{@{\extracolsep{\fill}} l r r r r r}
\toprule
\textbf{Metric} & \textbf{Observed} & \textbf{Null Mean} & \textbf{Null Std. Dev.} & \textbf{Z-Score} & \textbf{P-Value} \\
\midrule
Nodes ($N$) & 4,650 & 21.6 & 4.6 & 1,012.6 & $< 0.001$ \\
Edges ($E$) & 7,007 & 18.9 & 3.4 & 2,042.6 & $< 0.001$ \\
Connected Components & 1,313 & 8.9 & 2.1 & 630.9 & $< 0.001$ \\
Size of Largest Component & 40 & 4.3 & 1.0 & 36.1 & $< 0.001$ \\
Avg. Edge Weight & 0.904 & 0.892 & 0.012 & 1.0 & 0.30 \\
Mean Degree ($\langle k \rangle$) & 3.01 & 1.77 & 0.17 & 7.3 & $< 0.001$ \\
Multi-firm Insiders & 23 & 0.6 & 0.9 & 24.6 & $< 0.001$ \\
Prop. Multi-firm Insiders & $3.2 \times 10^{-4}$ & $0.1 \times 10^{-4}$ & $0.1 \times 10^{-4}$ & 22.4 & $< 0.001$ \\
Ultra-strong Ties & 3,472 & 7.9 & 1.6 & 2,199.1 & $< 0.001$ \\
\bottomrule
\end{tabular*}
\vspace{2mm}

\caption{\textbf{Detailed topological statistics for the Calibrated Generative Null Model.} This model anchors firm and insider activity to their empirical lifespans but randomizes specific trade dates. The immense Z-scores across structural metrics (Nodes, Edges, Components) confirm that the observed network structure is highly unlikely to arise from uncoordinated trading activity constrained only by tenure. P-values of 0.0 indicate $p < 10^{-100}$.}
\label{tab:supp_calibrated}
\end{table}

\begin{table}[h]
\centering
\setlength{\tabcolsep}{0pt} 
\begin{tabular*}{\textwidth}{@{\extracolsep{\fill}} l r r r r r}
\toprule
\textbf{Metric} & \textbf{Observed} & \textbf{Null Mean} & \textbf{Null Std. Dev.} & \textbf{Z-Score} & \textbf{P-Value} \\
\midrule
Nodes ($N$) & 4,650 & 776.2 & 22.8 & 169.7 & $< 0.001$ \\
Edges ($E$) & 7,007 & 1,446.1 & 31.0 & 179.3 & $< 0.001$ \\
Connected Components & 1,313 & 257.9 & 9.7 & 109.0 & $< 0.001$ \\
Size of Largest Component & 40 & 41.0 & 0.0 & $-\infty$ & $< 0.001$ \\
Avg. Edge Weight & 0.904 & 0.869 & 0.001 & 25.5 & $< 0.001$ \\
Mean Degree ($\langle k \rangle$) & 3.01 & 3.73 & 0.09 & -7.6 & $< 0.001$ \\
Multi-firm Insiders & 23 & 3.9 & 1.3 & 14.2 & $< 0.001$ \\
Prop. Multi-firm Insiders & $3.2 \times 10^{-4}$ & $0.6 \times 10^{-4}$ & $0.2 \times 10^{-4}$ & 14.2 & $< 0.001$ \\
Ultra-strong Ties & 3,472 & 340.7 & 23.7 & 132.0 & $< 0.001$ \\
\bottomrule
\end{tabular*}

\caption{\textbf{Detailed topological statistics for the Constrained Shuffled Null Model (Quarterly Stratification).} This model preserves the exact volume and timing of trades within each firm-quarter stratum but shuffles insider identities. The observed network remains significantly more structured (higher clustering, stronger ties) than this "environmental confounding" baseline. Note that the largest component size in the null model is highly deterministic (Std. Dev. $\approx 0$).}
\label{tab:supp_shuffled}
\end{table}

\newpage

\section{Underlying conditions for anomaly detection via OddBall in weighted graphs}\label{appendix:conditions}
\addcontentsline{supp}{section}{G\quad Underlying conditions for anomaly detection via OddBall in weighted graphs}

\begin{table}[h]
\centering
\caption{Verification of egonet power law conditions. Each condition represents a power-law relationship between egonet structural features. All four conditions are satisfied for this dataset.} \label{tab:egonet_conditions}
{\small
\begin{tabular}{@{}l c c c c@{}}
\toprule
\textbf{Condition} & \shortstack{\textbf{Variable} \\ \textbf{Relationship}} & \shortstack{\textbf{Fitted} \\ \textbf{Exponent}} & \shortstack{\textbf{Theoretical} \\ \textbf{Range}} & \shortstack{\textbf{Condition} \\ \textbf{Met?}} \\
\midrule
Egonet Density Power Law & $E_i \propto N_i^{\alpha}$ & $\alpha = 1.52$ & $1 \leq \alpha \leq 2$ & \checkmark \\
Egonet Weight Power Law & $W_i \propto E_i^{\beta}$ & $\beta = 1.21$ & $\beta \geq 1$ & \checkmark \\
Egonet Weighted Eigenvalue Power Law & $\lambda_{w,i} \propto W_i^{\gamma}$ & $\gamma = 0.75$ & $0.5 \leq \gamma \leq 1$ & \checkmark \\
Egonet Rank Power Law & $W_{i,j} \propto R_{i,j}^{\theta}$ & $\theta = -1.15$ & $\theta \leq 0$ & \checkmark \\
\bottomrule
\end{tabular}}
\end{table}

\begin{figure}[H]
\centering

\begin{subfigure}[b]{0.45\textwidth}
    \centering
    \includegraphics[width=\linewidth]{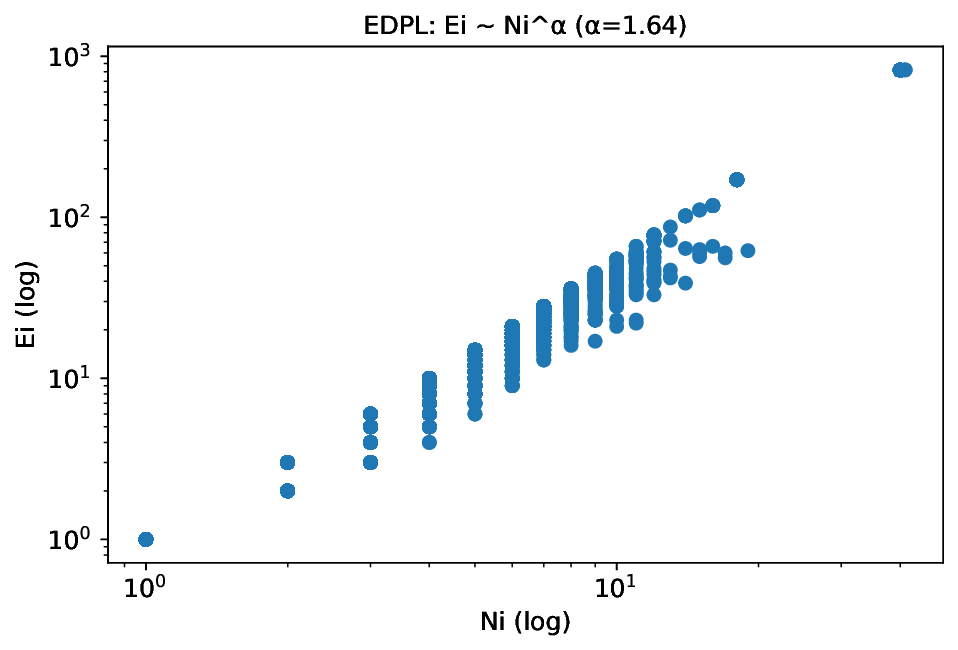}
    \caption{EDPL: $E_i \propto N_i^\alpha$}
    \label{fig:edpl}
\end{subfigure}
\hfill
\begin{subfigure}[b]{0.45\textwidth}
    \centering
    \includegraphics[width=\linewidth]{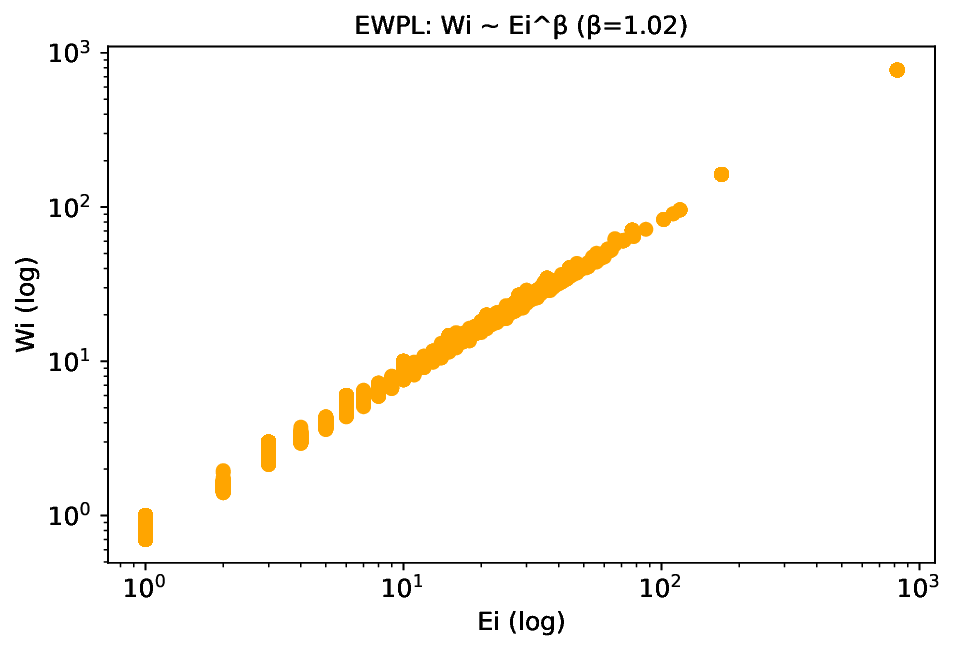}
    \caption{EWPL: $W_i \propto E_i^\beta$}
    \label{fig:ewpl}
\end{subfigure}

\vspace{0.5cm}

\begin{subfigure}[b]{0.45\textwidth}
    \centering
    \includegraphics[width=\linewidth]{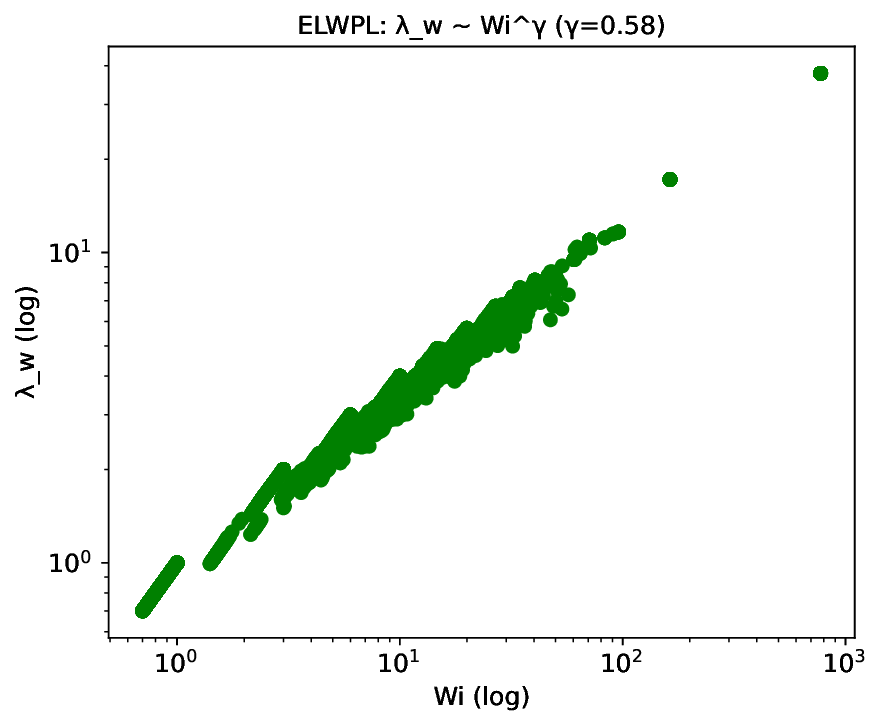}
    \caption{ELWPL: $\lambda_{w,i} \propto W_i^\gamma$}
    \label{fig:elwpl}
\end{subfigure}
\hfill
\begin{subfigure}[b]{0.45\textwidth}
    \centering
    \includegraphics[width=\linewidth]{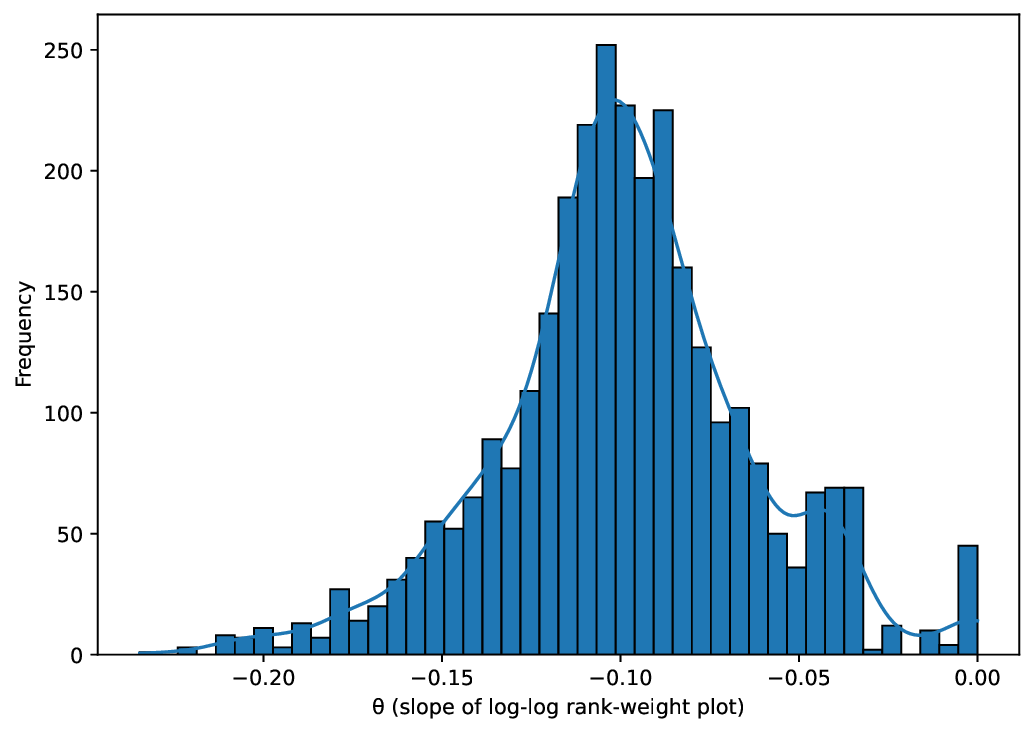}
    \caption{ERPL: $W_{i,j} \propto R_{i,j}^\theta$}
    \label{fig:erpl}
\end{subfigure}
\vspace{4mm} 
\caption{Verification of Egonet Power Law Conditions on Log-Log Plots. Each subplot displays the power-law fit for a different structural egonet property. The abbreviations correspond to the conditions listed in table 4.}
\label{fig:egonet-laws}
\end{figure}

\clearpage

\restoregeometry
\endgroup

%% file: sn-bibliography.bib
@article{afzali2021network,
  title={Network centrality and value relevance of insider trading: Evidence from Europe},
  author={Afzali, Mansoor and Martikainen, Minna},
  journal={Financial Review},
  volume={56},
  number={4},
  pages={793--819},
  year={2021},
  publisher={Wiley Online Library}
}

@article{el2021network,
  title={Network centrality, connections, and social capital: Evidence from CEO insider trading gains},
  author={El-Khatib, Rwan and Jandik, Dobrina and Jandik, Tomas},
  journal={Financial Review},
  volume={56},
  number={3},
  pages={433--457},
  year={2021},
  publisher={Wiley Online Library}
}

@article{el2015ceo,
  title={CEO network centrality and merger performance},
  author={El-Khatib, Rwan and Fogel, Kathy and Jandik, Tomas},
  journal={Journal of Financial Economics},
  volume={116},
  number={2},
  pages={349--382},
  year={2015},
  publisher={Elsevier}
}

@article{goergen2019insider,
  title={Insider trading and networked directors},
  author={Goergen, Marc and Renneboog, Luc and Zhao, Yang},
  journal={Journal of Corporate Finance},
  volume={56},
  pages={152--175},
  year={2019},
  publisher={Elsevier}
}

@article{ahern2017information,
  title={Information networks: Evidence from illegal insider trading tips},
  author={Ahern, Kenneth R},
  journal={Journal of Financial Economics},
  volume={125},
  number={1},
  pages={26--47},
  year={2017},
  publisher={Elsevier}
}

@article{cohen2012decoding,
  title={Decoding inside information},
  author={Cohen, Lauren and Malloy, Christopher and Pomorski, Lukasz},
  journal={The Journal of Finance},
  volume={67},
  number={3},
  pages={1009--1043},
  year={2012},
  publisher={Wiley Online Library}
}

@article{berkman2020inside,
  title={Inside the director network: When directors trade or hold inside, interlock, and unconnected stocks},
  author={Berkman, Henk and Koch, Paul and Westerholm, P Joakim},
  journal={Journal of Banking \& Finance},
  volume={118},
  pages={105892},
  year={2020},
  publisher={Elsevier}
}

@misc{kulkarni2017network,
  title={Network-based anomaly detection for insider trading},
  author={Kulkarni, Adarsh and Mani, Priya and Domeniconi, Carlotta},
  journal={arXiv preprint arXiv:1702.05809},
  year={2017}
}

@article{tamersoy2014large,
  title={Large-scale insider trading analysis: patterns and discoveries},
  author={Tamersoy, Acar and Khalil, Elias and Xie, Bo and Lenkey, Stephen L and Routledge, Bryan R and Chau, Duen Horng and Navathe, Shamkant B},
  journal={Social Network Analysis and Mining},
  volume={4},
  pages={1--17},
  year={2014},
  publisher={Springer}
}

@online{sec_insider_forms,
  title={Insider Transactions and Forms 3, 4, and 5},
  author={U.S. Securities and Exchange Commission},
  journal={Investor Bulletin},
  year={n.d.},
  publisher={U.S. Securities and Exchange Commission},
}

@inproceedings{akoglu2010oddball,
  title={Oddball: Spotting anomalies in weighted graphs},
  author={Akoglu, Leman and McGlohon, Mary and Faloutsos, Christos},
  booktitle={Advances in Knowledge Discovery and Data Mining: 14th Pacific-Asia Conference, PAKDD 2010, Hyderabad, India, June 21-24, 2010. Proceedings. Part II 14},
  pages={410--421},
  year={2010},
  organization={Springer}
}

@inproceedings{sun2024prediction,
  title={Prediction of Insider Trading Behaviours through Highly Interconnected Core People Network with Graph Machine Learning},
  author={Sun, Yinghuai and Gorduza, Dragos},
  booktitle={2024 6th International Conference on Robotics, Intelligent Control and Artificial Intelligence (RICAI)},
  pages={775--779},
  year={2024},
  organization={IEEE}
}

@article{augustin2019informed,
  title={Informed options trading prior to takeover announcements: Insider trading?},
  author={Augustin, Patrick and Brenner, Menachem and Subrahmanyam, Marti G},
  journal={Management Science},
  volume={65},
  number={12},
  pages={5697--5720},
  year={2019},
  publisher={INFORMS}
}

@article{cziraki2021dollar,
  title={The dollar profits to insider trading},
  author={Cziraki, Peter and Gider, Jasmin},
  journal={Review of Finance},
  volume={25},
  number={5},
  pages={1547--1580},
  year={2021},
  publisher={Oxford University Press}
}

@article{gomez2024effect,
  title={The effect of mandatory disclosure dissemination on information asymmetry among investors: Evidence from the implementation of the EDGAR system},
  author={Gomez, Enrique A},
  journal={The Accounting Review},
  volume={99},
  number={1},
  pages={235--257},
  year={2024},
  publisher={American Accounting Association}
}

@incollection{fidrmuc2008insider,
  title={Insider trading, news releases, and ownership concentration},
  author={Fidrmuc, Jana and Goergen, Marc and Renneboog, Luc},
  booktitle={Insider Trading},
  pages={309--370},
  year={2008},
  publisher={CRC Press}
}

@misc{guay2021insider,
  title={Determinants of Insider Trading Windows},
  author={Guay, Wayne R. and Kim, Shawn and Tsui, David},
  journal={Harvard Law School Forum on Corporate Governance},
  year={2021},
  publisher={Harvard Law School},
}

@article{gauvin2022randomized,
  title={Randomized reference models for temporal networks},
  author={Gauvin, Laetitia and G{\'e}nois, Mathieu and Karsai, M{\'a}rton and Kivelä, Mikko and Takaguchi, Taro and Valdano, Eugenio and Vestergaard, Christian L},
  journal={SIAM Review},
  volume={64},
  number={4},
  pages={763--830},
  year={2022},
  publisher={SIAM}
}

@article{fagiolo2013null,
  title={Null models of economic networks: the case of the world trade web},
  author={Fagiolo, Giorgio and Squartini, Tiziano and Garlaschelli, Diego},
  journal={Journal of economic interaction and coordination},
  volume={8},
  number={1},
  pages={75--107},
  year={2013},
  publisher={Springer}
}

@article{crane2019institutional,
  title={Institutional investor cliques and governance},
  author={Crane, Alan D and Koch, Andrew and Michenaud, S{\'e}bastien},
  journal={Journal of Financial Economics},
  volume={133},
  number={1},
  pages={175--197},
  year={2019},
  publisher={Elsevier}
}

@article{casiraghi2019block,
  title={The block-constrained configuration model},
  author={Casiraghi, Giona},
  journal={Applied Network Science},
  volume={4},
  number={1},
  pages={123},
  year={2019},
  publisher={Springer}
}

@article{gauvin2013activity,
  title={Activity clocks: spreading dynamics on temporal networks of human contact},
  author={Gauvin, Laetitia and Panisson, Andr{\'e} and Cattuto, Ciro and Barrat, Alain},
  journal={Scientific reports},
  volume={3},
  number={1},
  pages={3099},
  year={2013},
  publisher={Nature Publishing Group UK London}
}

@online{jones2024insidertrading,
  author       = {Jones, Michael T. and Magee, Jessica B. and Kernisky, Allison},
  title        = {3 French Hens? No. SEC Presses Enforcement on Insider Trading, Shadow Trading, Reg FD},
  year         = {2024},
  month        = dec,
  date         = {2024-12-12},
  note         = {Holland \& Knight SEC{O}nd Opinions Blog, Season's Readings Series}
}

@article{bhattacharya2014insider,
  title={Insider trading controversies: A literature review},
  author={Bhattacharya, Utpal},
  journal={Annu. Rev. Financ. Econ.},
  volume={6},
  number={1},
  pages={385--403},
  year={2014},
  publisher={Annual Reviews}
}

@article{fishman1992insider,
  title={Insider trading and the efficiency of stock prices},
  author={Fishman, Michael J and Hagerty, Kathleen M},
  journal={The RAND Journal of Economics},
  pages={106--122},
  year={1992},
  publisher={JSTOR}
}

@article{arif2022audit,
  title={Audit process, private information, and insider trading},
  author={Arif, Salman and Kepler, John D and Schroeder, Joseph and Taylor, Daniel},
  journal={Review of Accounting Studies},
  volume={27},
  number={3},
  pages={1125--1156},
  year={2022},
  publisher={Springer}
}

@article{priyadarshi2024comprehensive,
  title={A comprehensive review on insider trading detection using artificial intelligence},
  author={Priyadarshi, Prashant and Kumar, Prabhat},
  journal={Journal of Computational Social Science},
  volume={7},
  number={2},
  pages={1645--1664},
  year={2024},
  publisher={Springer}
}

@article{li2022internet,
  title={Internet financial fraud detection based on graph learning},
  author={Li, Ranran and Liu, Zhaowei and Ma, Yuanqing and Yang, Dong and Sun, Shuaijie},
  journal={IEEE Transactions on Computational Social Systems},
  volume={10},
  number={3},
  pages={1394--1401},
  year={2022},
  publisher={IEEE}
}

@article{oliveira2025complex,
  title={Complex networks-based anomaly detection for financial transactions in anti-money laundering},
  author={Oliveira, Rodrigo Marcel Araujo and Sant’Anna, Angelo Marcio Oliveira and Ferreira, Paulo Henrique},
  journal={Forensic Science International: Digital Investigation},
  volume={55},
  pages={302005},
  year={2025},
  publisher={Elsevier}
}

@inproceedings{basu2018novel,
  title={A novel graph analytic approach to monitor terrorist networks},
  author={Basu, Kaustav and Zhou, Chenyang and Sen, Arunabha and Goliber, Victoria Horan},
  booktitle={2018 IEEE Intl Conf on Parallel \& Distributed Processing with Applications, Ubiquitous Computing \& Communications, Big Data \& Cloud Computing, Social Computing \& Networking, Sustainable Computing \& Communications (ISPA/IUCC/BDCloud/SocialCom/SustainCom)},
  pages={1159--1166},
  year={2018},
  organization={IEEE}
}

@article{cline2017persistence,
  title={The persistence of opportunistic insider trading},
  author={Cline, Brandon N and Gokkaya, Sinan and Liu, Xi},
  journal={Financial Management},
  volume={46},
  number={4},
  pages={919--964},
  year={2017},
  publisher={Wiley Online Library}
}

@article{yeung2025garbage,
  title={Garbage in garbage out? Impacts of data quality on criminal network intervention},
  author={Yeung, Wang Ngai and Di Clemente, Riccardo and Lambiotte, Renaud},
  journal={EPJ Data Science},
  volume={14},
  number={1},
  pages={37},
  year={2025},
  publisher={Springer}
}

@article{squartini2011analytical,
  title={Analytical maximum-likelihood method to detect patterns in real networks},
  author={Squartini, Tiziano and Garlaschelli, Diego},
  journal={New Journal of Physics},
  volume={13},
  number={8},
  pages={083001},
  year={2011},
  publisher={IOP Publishing}
}

@article{fried2013insider,
  title={Insider trading via the corporation},
  author={Fried, Jesse M},
  journal={U. Pa. L. Rev.},
  volume={162},
  pages={801},
  year={2013},
  publisher={HeinOnline}
}

@misc{isc2019merger,
  author = {{International Speedway Corporation}},
  title  = {International Speedway Corporation Announces Merger Agreement with NASCAR Holdings, Inc.},
  year   = {2019},
  howpublished = {\url{sec.gov/edgar}}
}

@misc{plug2022shareholderalert,
  author       = {{Schubert Jonckheer \& Kolbe LLP}},
  title        = {Shareholder Alert: Plug Power Inc. (PLUG) Officers and Directors Under Investigation for Possible False Statements and Insider Trading},
  year         = {2022},
  note         = {Press release, PR Newswire},
  howpublished = {\url{https://www.prnewswire.com/news-releases/shareholder-alert-plug-power-inc}}
}

@misc{sec2018aveo,
  author = {{U.S. Securities and Exchange Commission}},
  title  = {SEC Obtains Final Judgments Against Former AVEO Pharmaceuticals Executives},
  year   = {2018},
  url    = {https://www.sec.gov/enforcement-litigation/litigation-releases/lr-24062}
}
